\documentclass[useAMS,usenatbib]{mnras}

\usepackage{
amsmath,
amssymb,
graphicx,
xspace,
hyperref,
xcolor,
}

\newcommand\citedummy[1]{}
\newcommand\eg{e.\,g.,\xspace}
\newcommand\ie{i.\,e.,\xspace}

\newcommand\spose[1]{\hbox to 0pt{#1\hss}}
\newcommand\lta{\mathrel{\spose{\lower 3pt\hbox{$\mathchar"218$}}
     \raise 2.0pt\hbox{$\mathchar"13C$}}}
\newcommand\gta{\mathrel{\spose{\lower 3pt\hbox{$\mathchar"218$}}
     \raise 2.0pt\hbox{$\mathchar"13E$}}}

\newcommand\vect[1]{\ensuremath{\boldsymbol{#1}}}
\renewcommand{\d}[1]{\ensuremath{\operatorname{d}\!{#1}}}

\newcommand\jz{\ensuremath{j}\xspace}
\newcommand\rexp{\ensuremath{R_\mathrm{exp}}\xspace}
\newcommand\mlr{\ensuremath{\displaystyle\Upsilon}\xspace}

\newcommand\vphi{\vcirc}
\newcommand\vcirc{\ensuremath{v_\mathrm{c}}\xspace}

\newcommand\jmean{\ensuremath{\langle j\rangle}\xspace}
\newcommand\chitwo{\ensuremath{\chi^2}\xspace}
\newcommand\changed[1]{{#1}}

\newcommand\half{{\textstyle{\frac{1}{2}}}}

\title[Maximum entropy discs]{Galactic disc profiles and a
    universal angular momentum distribution from statistical physics}
\author[Herpich et al.]{
    Jakob Herpich$^{1}$%
\thanks{E-mail: herpich@mpia.de}%
\thanks{Member of the International Max Planck Research School for
    Astronomy and Cosmic Physics at the University of Heidelberg,
    IMPRS-HD, Germany.
    },
    Scott Tremaine$^{2}$,
    Hans-Walter Rix$^{1}$
    \\
    $^{1}$Max-Planck-Institut f\"ur Astronomie, K\"onigstuhl 17, D-69117 Heidelberg, Germany\\
    $^{2}$Institute for Advanced Study, Princeton, NJ 08540, USA
}
\begin{document}

\maketitle

\begin{abstract}
We show that the stellar surface brightness profiles
in disc galaxies---observed to be approximately exponential---can be explained
if radial migration efficiently scrambles the individual stars' angular momenta
while conserving the circularity of the orbits and the total mass and angular momentum.
In this case, the disc's distribution of
specific angular momenta $j$ should be near a maximum entropy state and therefore
approximately exponential, $\d{N} \propto \exp(-j/\jmean) \d{j}$.
This distribution translates to a surface density profile that is generally not an
exponential function of radius:
$\Sigma(R) \propto \exp\left[-R/R_e({\scriptstyle{R}})\right]/
\left (R R_e({\scriptstyle{R}})\right)\left(
1+\d{\,\log\vphi(R)}/\d{\,\log R}\right),$ for
a rotation curve $\vphi(R)$ and 
$R_e({\scriptstyle{R}})\equiv \jmean/\vphi(R)$.
We show that such a profile matches the 
observed surface brightness profiles of disc-dominated galaxies as well as the empirical exponential profile.   
Disc galaxies that exhibit population gradients cannot have fully reached a
maximum entropy state but appear to be close enough that their surface brightness profiles
are well fit by this idealized model.
\end{abstract}


\section{Introduction}
\label{sec:intro}

The observed radial surface brightness (or surface density) profiles of stars in disc galaxies are approximately exponential,
$\Sigma_\star(R)\propto \exp(-R/R_\mathrm{exp})$. This result dates back to work by \citet{Patterson1940} and \citet{deVaucouleurs1959} but was extended and popularized by \citet{Freeman1970}, and we will use the term ``Freeman discs" to describe galaxies of this kind.
In some galaxies there are clear systematic deviations
 from the Freeman disc at large radii \citep{Pohlen2002,Erwin2005,Pohlen2006}; 
and at small radii there are commonly more stars than expected in a Freeman disc, an excess usually attributed to a bulge component. 
Nonetheless, the Freeman disc has proven to be a 
remarkably good approximation to the stellar profile in a large majority of disc-dominated galaxies. 

Attempts have been made to explain these profiles as a reflection of the angular momentum 
distribution of the gas from which the stars are formed 
\citep[\eg][]{Fall1980,Dalcanton1997,Dutton2009}.
Here we set out to provide a qualitatively different explanation,
one that is based on the dynamical evolution of the disc rather than the initial conditions, a possibility that
has not received much attention in the literature until recently  \citep{Lin1987,Yoshii1989,Ferguson2001,Elmegreen2013,Elmegreen2016,Struck2017}.
In this paper, we argue on the basis of a simple analytical model 
that disc surface density profiles are a 
natural consequence of the equilibrium statistical mechanics of galactic discs. 

While the characteristic size of a stellar disc presumably reflects its initial angular
momentum, a wide range of work suggests that the present-day \emph{profile} of the disc is not primarily a reflection of initial conditions.
Simulations of the formation of disc galaxies find that the profile of stellar 
birth radii is quite often far from a Freeman disc  \citep{Debattista2006,Roskar2008,Roskar2012,Minchev2012,Berrier2015,Herpich2015,Herpich2016}.
Nevertheless, simulations of both isolated disc galaxies
 \citep{Steinmetz1995,Roskar2008,Roskar2008a,Roskar2012,Minchev2012,Elmegreen2013,Berrier2015,Herpich2015,Herpich2016},
and galaxies forming in a cosmological context \citep{Katz1992,Navarro1994,Sommer-Larsen1999,Sommer-Larsen2003,Governato2004,Governato2007,Robertson2004,Okamoto2005,Guedes2011,Stinson2013}
show that approximate Freeman disc profiles eventually emerge. 

This evolution arises because stars do not remain on the orbits on which they were born, even in the absence of substantive mergers.
Both analytic arguments and numerical experiments show that
the angular momenta of individual disc particles are 
``churned'' or ``shuffled'' by corotating, transient, non-axisymmetric perturbations such as 
spiral arms \citep[and references therein]{Sellwood2014}.
This mechanism preserves the small orbital eccentricities of the initial disc \citep[\eg][]{Sellwood2002,Roskar2008,Roskar2012},
can be very efficient \citep{Sellwood2002,Roskar2012} and does not change the total angular momentum of
the disc stars appreciably (see Section 2.3.2); it has been dubbed \emph{radial migration}.

In this paper we explore which radial surface density profiles
are the end state of radial migration. We adopt the idealizing,
but we believe sensible, approximations that stars are born, and remain, on 
circular orbits, and that the total angular momentum and mass of the disc
are conserved, while radial migration scrambles the individual stellar
angular momenta. If the angular momenta are completely scrambled, then their final distribution should have the maximum entropy possible,  
subject to the just-mentioned constraints.

Applying concepts from statistical mechanics to describe the
equilibrium phase-space distribution in a self-gravitating stellar
system has proven problematic for a variety of reasons, the most
important of which is that the volume of phase space enclosed by a
surface of constant energy is not finite \citep[see][Section 7.3]{Binney2008}.
However, we will argue that the
angular momentum distribution resulting from radial migration \emph{can} be
described by maximizing a suitably defined entropy, using methods
that are completely analogous to deriving the Boltzmann distribution
for ideal gases. 

The paper is structured as follows. In Section \ref{sec:toy_model} we describe our model and
in Section \ref{sec:application} we
derive an analytic expression for the surface density profile
corresponding to a given galactic rotation curve.
In Section \ref{sec:validation} we compare the predictions of this work to data from observations
and simulations.
Finally, we discuss the results and implications of our model in Section \ref{sec:discussion}.
In Appendix \ref{sec:self_consistent} we derive the unique maximum entropy
surface density profile in the special case of a disc that has no surrounding dark halo. 

\section{An analytic model for radial profiles of disc galaxies}
\label{sec:toy_model}

\subsection{Choice of phase-space variables}
\label{sec:phase_space}

We work in cylindrical coordinates $\vect r=(R,\varphi,z)$ and use action-angle 
variables $(\vect J, \vect\theta)=(J_R,J_\varphi,J_z; \theta_R,\theta_\varphi,\theta_z)$ as phase-space coordinates
\citep{Binney2008}. Action-angle variables are
canonical so
$\d{\vect\theta}\d{\vect J}=\d{\vect r}\d{\vect p}$ where $\vect p$ is
the ordinary Cartesian momentum conjugate to $\vect r$. 
Since stars are born on
nearly circular, coplanar orbits and migration does not excite
eccentricities or inclinations (see \S\ref{sec:intro}) we may assume that $J_R=J_z=0$. The azimuthal action $J_\varphi$ is equal to the angular
momentum per unit mass along the $z$-axis, $j$, and we can restrict ourselves to the two-dimensional
manifold in phase space $(J_\varphi,\theta_\varphi)$. The angle
variables $\theta_\varphi$ are uniformly distributed so the state of a
disc of $N$ stars is fully specified by the angular momenta
$(j_1,\ldots,j_N)$. Since $N\gg1$ and the stars interact weakly, the state of the
disc is also fully described by the one-particle distribution function
$F(j)$, where $F(j)\d j \d\theta_\varphi$ is the number of stars
in a phase-space element $\d j \d\theta_\varphi$. 

Since migration is driven by transient spiral arms and other
structures much more massive than stars, there is no mass segregation
during migration and with no loss of generality we can assume that all
stars have the same mass $m$. 

\subsection{Entropy}
\label{sec:entropy}

The central premise of this paper is that radial migration 
scrambles the individual angular momenta of disc stars. Ultimately, this will
result in a distribution of stars in
the $N$-dimensional space with coordinates $(j_1,\ldots,j_N)$ that is
uniform on the manifold allowed by the conserved quantities
(the total number of stars and the total angular momentum; see
\S\ref{sec:cons}). This is simply the ergodic hypothesis
that underlies the construction of the microcanonical ensemble in classical statistical
mechanics. Then standard arguments show
that if $N\gg 1$ and the interactions are weak, the relative probability
$\Omega$ associated with the distribution function $F(j)$ is given by
\begin{equation} 
{\mathfrak S}=-\int \d j\d\theta_\varphi\, F(j)\log
  F(j) = -2\pi \int \d j\, F(j)\log F(j)
\label{eq:entropy}
\end{equation}
is the Boltzmann entropy, and we have set Boltzmann's constant to unity. 

Applying equilibrium statistical mechanics to this problem 
implicitly assumes that radial migration is efficient enough to completely scramble stellar orbits across
the whole radial range of the disc, and does so on a time-scale shorter
compared to the age of the galaxy. We caution that this
premise cannot be strictly true, since that would 
produce a stellar disc without abundance or age gradients.

\subsection{Conserved Quantities}
\label{sec:cons}

\subsubsection{Total number of stars}
 We assume that the number $N$ of stars in the
disc is fixed; thus
\begin{equation}
N= 2\pi\! \int \d j\, F(j)= \mbox{constant}.
\label{eq:ndef}
\end{equation}
In fact ongoing star formation is present in almost all discs that
exhibit the spiral structure that drives migration. However, focusing
our attention on the distribution of older stars (e.g., by measuring the surface brightness profile at red or infrared wavelengths), we can minimize
contamination by recent star formation.

\subsubsection{Total angular momentum}

\label{sec:angmom}
We also assume that the total angular
momentum $J$ of the disc is fixed, thus
\begin{equation}
J=Nm\jmean= 2\pi m\int \d j\, jF(j)= \mbox{constant}.
\label{eq:jdef}
\end{equation}
The validity of this assumption may be limited by several processes.

A central bar can exert torques on the dark halo that drain angular
momentum from the disc. Such torques can be substantial but
are difficult to estimate because the dark halo density near the
centre of the galaxy is poorly determined (see \citealt{Sellwood2014}
for a review). Typically, however, the effects of bar--halo
torques are likely to be small since (i) the angular momentum of the
bar is much less than that of the disc, because it has only a small fraction of the disc mass and a
smaller radius of gyration; (ii) most bars have high pattern speeds
\citep{Sellwood2014}, which suggests that they have not lost much of
their original angular momentum.

Spiral structure can also transfer angular momentum from the disc to the halo
\citep{Mark1976,Fuchs2004} but for plausible values of the pitch angle
and amplitude of the spirals the torques are too small to modify the
disc angular momentum significantly. This would be true even if (unrealistically) the strong ``grand-design" spiral
patterns seen in some galaxies survived unchanged for a Hubble time \citep{TO1999}.

Galaxy discs continuously build up angular momentum as they grow by accretion.
This process violates our assumptions of mass and angular momentum conservation.
However, as long as radial migration is effective on a time-scale shorter 
than that of the total mass and angular momentum change due to late-epoch gas infall, 
the approximations described here remain sensible.

In \S\ref{sec:discussion} we will discuss the
validity of these assumptions in more detail.

\subsubsection{Energy} The total potential energy of the disc can be written as
\begin{align}
U &= 2\pi\!\int \d R\, R\Sigma(R) \Phi_h(R) \nonumber \\
&\quad + 2\pi^2 \!\int \d R \d R'\, RR'\Sigma(R)\Sigma(R')W(R,R'),
\end{align}
here $\Sigma(R)$ is the surface density of the disc at radius $R$,
$\Phi_h(R)$ is the gravitational potential due to the dark halo and
the kernel $W(R,R')$ is the gravitational potential between two
coplanar rings of unit mass at $R$ and $R'$; two expressions for this kernel are 
\begin{align} \label{eq:kernel}
W(R,R')&=-\frac{2G}{\pi(R+R')}K\left(\frac{2\sqrt{RR'}}{R+R'}\right),
  \nonumber \\
&=-\frac{2G}{\pi R_>}K\left(\frac{R_<}{R_>}\right),
\end{align}
where $K$ is a complete elliptic integral and $R_<$ and $R_>$ are, respectively, the smaller and larger of $R$
and $R'$. The gravitational potential in the disc plane is
\begin{equation}\label{eq:pot}
\Phi(R)=\Phi_h(R)+ 2\pi\int \d  R' R'\Sigma(R')W(R,R')
\end{equation}
and the kinetic energy of the disc is
\begin{equation}
T=\pi\int \d R\,\Sigma(R) R^2\frac{\d\Phi}{\d R}.
\end{equation}
The surface density is
related to the distribution function by
\begin{align}
\Sigma(R)&=\frac{m}{R}F(j)\frac{\d j}{\d R},
\label{eq:surface_density_ang_mom_distribution} \\
j^2(R)&=R^3\frac{\d \Phi}{\d R}.
\label{eq:ang_mom_rot_curve}
\end{align}

We have chosen not to include a constraint so
that the total disc energy is conserved, for two reasons. 
First, spiral structure requires gas in the disc and hence dissipation;
without dissipation, transient spirals heat the disc and quench the
formation of further spirals. Thus we do not expect the energy of
migrating discs to be conserved.  
Second, an energy constraint would add an extra free parameter to our models and our intent is to explore the simplest possible models for maximum entropy discs. 

Nevertheless, the  disc's total energy does play an important role.
Maximum entropy discs are a possible end state of migration only if they
have lower energy than other discs with the same mass and angular momentum. We show in Appendix \ref{sec:h_theorem} that this is likely to be the case. 

\subsection{The maximum entropy disc}

The maximum entropy state consistent with a fixed number of stars $N$ and
fixed total angular momentum $J$ is determined by the variational
equation:
\begin{align}
0=&\delta {\mathfrak S} - \alpha \delta N - \beta\delta J\nonumber \\
=& -2\pi\int \d
j\,\delta F(j)\big[(1+\log F(j)) +\alpha +\beta j\Big], 
\end{align}
where $\alpha$ and $\beta$ are Lagrange multipliers. The
solution to this equation is $F(j)=\exp(-1-\alpha-\beta j)$. 
Substituting back into equations \eqref{eq:ndef} and \eqref{eq:jdef},
it is straightforward to determine $\alpha$ and $\beta$ and rewrite the
distribution function as
\begin{equation} 
F(j)=\frac{N}{2\pi \jmean}\exp\big(-j/\jmean\big).
\label{eq:dfdefa}
\end{equation}
The fraction of stars with angular momentum less than $j$ is
\begin{equation}
\frac{N(<j)}{N}=1-\exp\big(-j/\jmean\big).
\label{eq:intdist}
\end{equation}
Equation \eqref{eq:dfdefa} or \eqref{eq:intdist} encapsulates a remarkable
result: given our assumptions, the maximum entropy distribution of
specific angular momentum of a stellar disc 
is always exponential, independent of the actual gravitational potential that determines the rotation
curve.  We now translate this
distribution into surface density profiles.

\section{Application to galaxy discs}
\label{sec:application}

To compare the maximum entropy distribution function \eqref{eq:dfdefa} to observations 
for a given rotation curve $\vphi(R)$, we need to relate the surface density $\Sigma(R)$ to the distribution function $F(j)$. Circular orbits are stable if and only if the angular momentum increases with radius, so 
$\vphi(R)$ defines a bijective mapping between
\jz and $R$ through the equation
\begin{equation}
    \jz(R) = R\vphi(R).
    \label{eq:angmom_radius}
\end{equation}
Moreover rotation curves are generally smooth \citep{Berrier2015} so the mapping is well-behaved. 

We can then compute the surface density profile from equations
(\ref{eq:surface_density_ang_mom_distribution}) and (\ref{eq:dfdefa}), 
\begin{equation}
        \Sigma(R)= \frac{mN}{2\pi R\jmean}\exp\left[-\frac{R\vphi(R)}\jmean\right]\left[\vphi(R)+R\frac{\d{\vphi}}{\d R}\right].
    \label{eq:surface_density}
\end{equation}
An alternative form is 
\begin{equation}
        \Sigma(R)= \frac{M}{2\pi R R_e(R)}\exp\left[-\frac{R}{R_e(R)}\right]\left(1+\frac{\d{\,\log\vphi}}{\d{\,\log R}}\right),
    \label{eq:surface_density2}
\end{equation}
with $R_e(R)\equiv \jmean / \vphi(R)$ and $M\equiv Nm$ the total mass of the disc. 

For a flat rotation curve, $\vphi(R)=\,$constant, $R_e$ is independent of radius and we have 
\begin{equation}
        \Sigma(R)= \frac{M}{2\pi R_e^2}\cdot \frac{R_e}{R}~\exp{\left(-\frac{R}{R_e}\right )}.
    \label{eq:flat_rot_profile}
\end{equation}
The surface density is an approximately exponential function of radius for $R\gg R_e$, and is $\propto R^{-1}$ for $R\ll R_e$. Alternatively, we may ask for which rotation curve 
the maximum entropy angular momentum distribution (\ref{eq:dfdefa}) has a Freeman  surface density distribution, $\Sigma_\star(R)\propto \exp(-R/R_\mathrm{exp})$. It is straightforward to show that in this case:
\begin{equation}
\vphi(R)=v_\infty\left[1-\frac{R_\mathrm{exp}}{R}\log(1+R/R_\mathrm{exp})\right],
\end{equation}
where $v_\infty\equiv \jmean/R_\mathrm{exp}$ is the circular speed at large radii. 

In this analysis we have derived the maximum entropy surface density distribution for a given gravitational potential or rotation curve arising from some combination of the disc and dark halo mass. 
In Appendix \ref{sec:self_consistent} we describe the special case in which there is no halo so the potential is derived self-consistently from the disc mass. In this case the surface density varies as $\exp[-(R/R_0)^{1/2}]$ at large radii and as $R^{-1}$ at small radii. In principle one could calculate self consistently the disc surface density distribution the halo density distribution, and the rotation curve resulting from efficient migration, but this is much more difficult, and the result would depend on the initial state of the disc and halo (see discussion in \S\ref{sec:discussion}).

\section{Validation of the model}
\label{sec:validation}

\subsection{Observational tests}
\label{sec:obserational_test}

We now ask whether observed surface density profiles of disc galaxies match the prediction from equation \eqref{eq:dfdefa} or 
\eqref{eq:surface_density2}. To do so we consider the circular velocity curve, or rotation curve, $\vphi(R)$ and the surface brightness profile $I(R)$ for a set of disc-dominated
galaxies. We assume that the stellar mass-to-light ratio \mlr is independent of radius so $I(R)\propto\Sigma(R)$, although the actual value of \mlr is not needed. This assumption is reasonable if observations at red (optical) bands are available, as population and dust extinction 
gradients have only modest impact \citep{vanderKruit_Freeman2011}.  

We use a set of 304 disc galaxies with $r$-band surface brightness profiles and H$\alpha$ rotation curves from \citet{Courteau1996,Courteau1997}\footnote{The data are available at
\url{http://www.cadc-ccda.hia-iha.nrc-cnrc.gc.ca/COURTEAU/}. Some galaxies have more than a single data set, in which case we choose the first one.}.
\citet{Courteau1997} also provides fitting functions\footnote{We use Courteau's Model 2, which fits the rotation curve to the form
$\vcirc(R)=v_c(1+R_t/R)^\beta/(1+R_t^\gamma/R^\gamma)^{1/\gamma}$.
There are four fitting parameters, $v_c$, $r_t$, $\gamma$, and $\beta$, but $\beta$ is set to $0$ for most of the galaxies. See the paper for more detail.}  and best-fitting parameters for the rotation curves, which we use to derive a smooth approximation to $\vphi(R)$.

 Based on these data, and our assumption of circular orbits,
we can calculate $F(j)$ using equations \eqref{eq:surface_density_ang_mom_distribution} and \eqref{eq:angmom_radius}. We then fit $\log F(j)$ to the log of the exponential form \eqref{eq:dfdefa} using a set of 60 equally spaced angular momenta $\left\{j_k\right\}$ and assuming equal weights for each point. The free parameters determined by the fit are the normalization and the mean angular momentum $\langle j\rangle$.
For the fit we use equal weights for every data point.

\begin{figure}
    \includegraphics[width=\columnwidth]{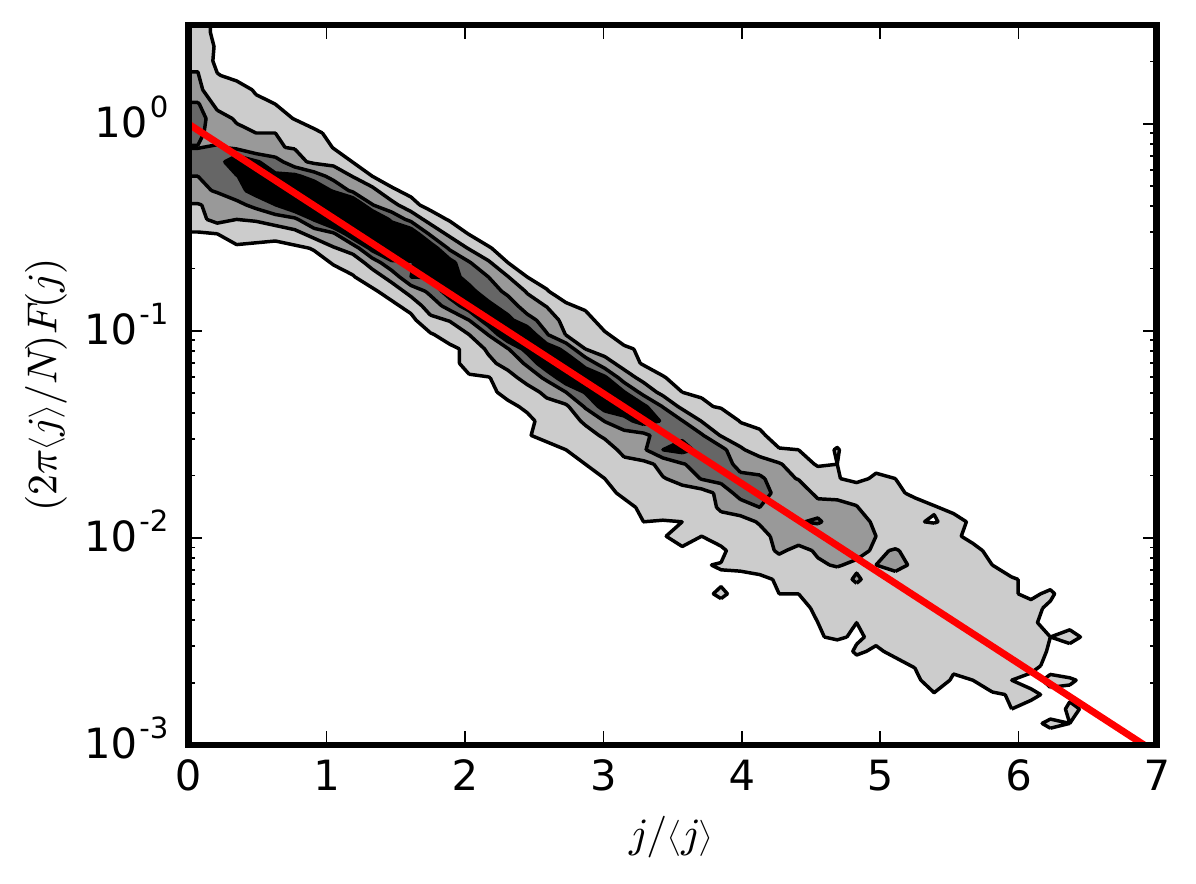}
    \caption{The distribution of specific angular momentum stacked
        for 304 disc galaxies from \citet{Courteau1996,Courteau1997}.
        Every distribution has been rescaled using two fit parameters (normalization and \jmean).
        The contours enclose 30, 50, 70 and 90\% of the plotted data points.
        The red line is the maximum entropy distribution  (equation  \ref{eq:dfdefa}).
        Despite some small but systematic deviations for small $j$ there is good agreement
        with the model prediction. 
    }
    \label{fig:ang_mom_hist}
\end{figure}

\begin{figure*}
    \includegraphics[width=.95\textwidth]{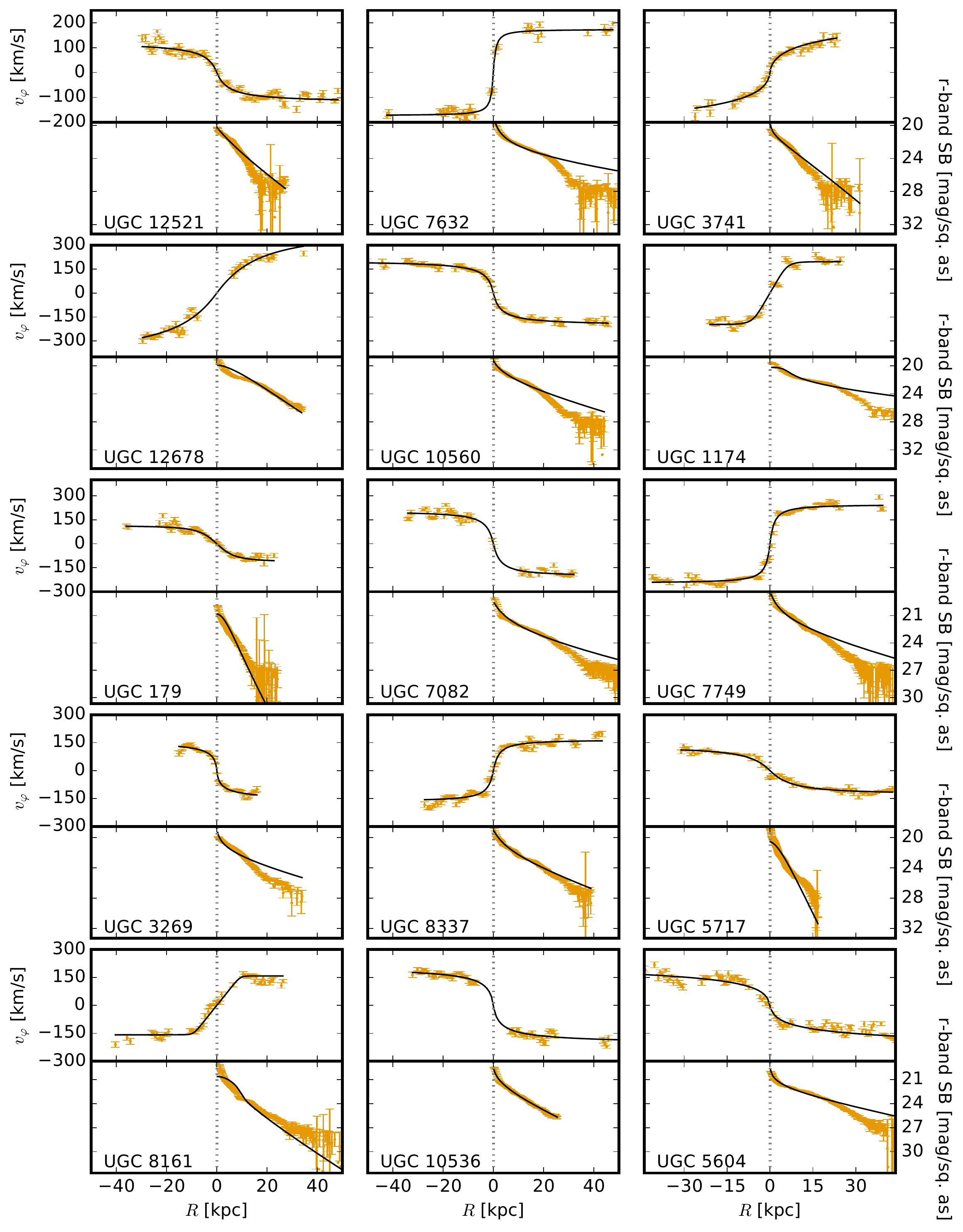}
    \caption{Model fits for 15  randomly selected galaxies from \citet{Courteau1996,Courteau1997}.
        For each galaxy the top panel shows the measured rotation curve (orange) and the
        Model 2 fit from \citet{Courteau1997} (black line).
        The bottom panel shows the measured surface brightness profile (orange) and the
        best-fit model profile for a maximum entropy disc (black line) derived from the
        rotation curve using equation \eqref{eq:surface_density} and the assumption of
        constant mass-to-light ratio \mlr.
        Distances are obtained from the redshifts assuming a Hubble constant
        of $H_0=70 \mathrm{\,km\ s}^{-1}\,\mbox{Mpc}^{-1}$.
        See text for a description of the fitting procedure.
    }
    \label{fig:courteau_fit}
\end{figure*}

Fig. \ref{fig:ang_mom_hist} shows the stacked specific angular momentum distributions of all the galaxies in our sample, in the normalized coordinates $j/\langle j\rangle$ and $2\pi\langle j\rangle F(j)/N$. 
The red straight line represents the prediction for maximum entropy discs, equation \eqref{eq:dfdefa}.
Broadly speaking the angular momentum profiles reconstructed from the data
are consistent with the predictions over much of the $j$-range. 
Only for  $j\lta\jmean$ there are some systematic deviations from the prediction; these could arise from contamination by bulge components or because our model assumption of circular orbits breaks down at small radii.
Note also that a number of \emph{individual} galaxies are clearly not consistent with our model, while others show almost perfectly exponential specific angular momentum distributions---this is seen more clearly in Fig. \ref{fig:courteau_fit}.  These results suggest that more factors than radial migration play a role in establishing the angular momentum profile, at least in some galaxies.

\begin{figure}
    \includegraphics[width=\columnwidth]{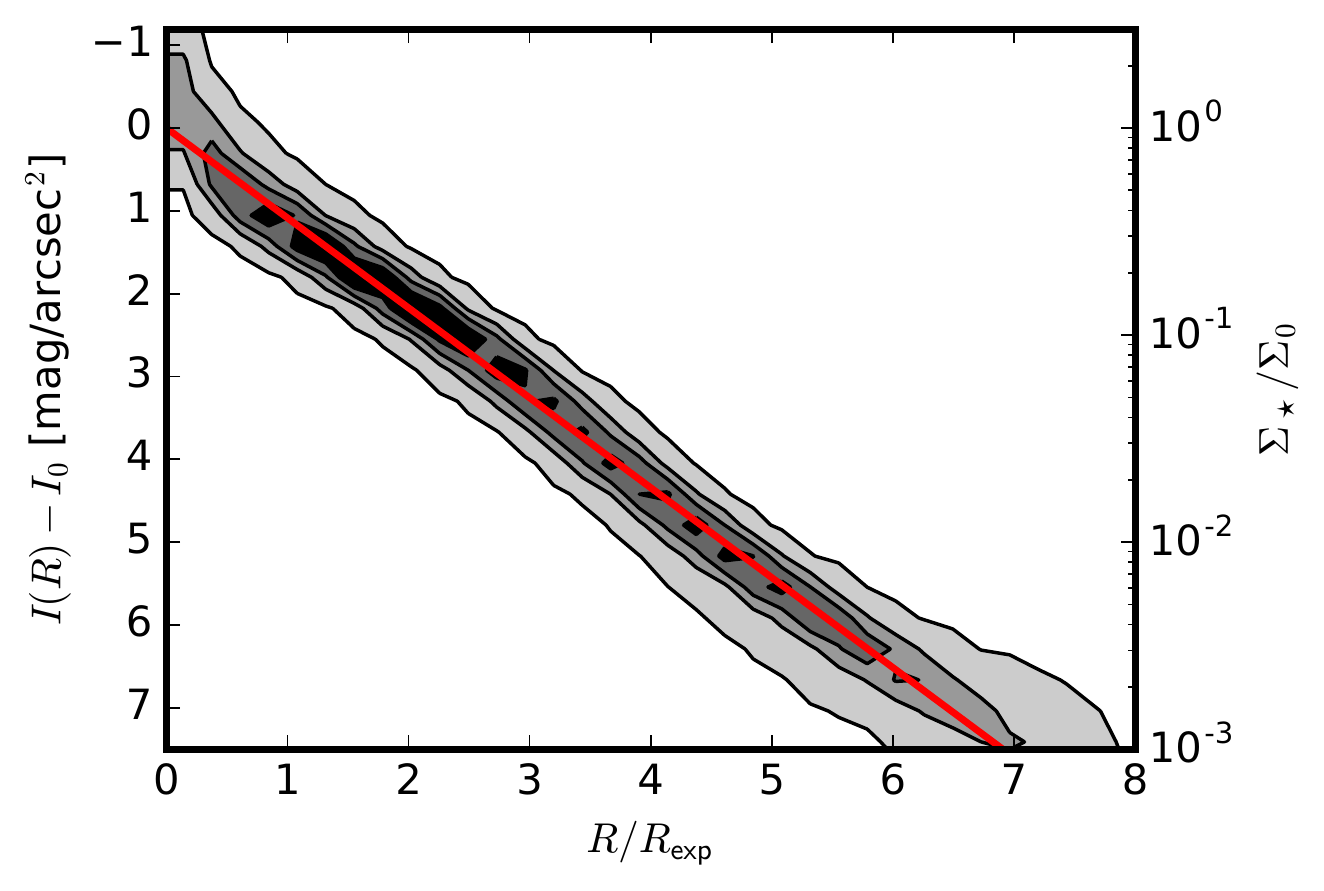}
    \caption{Stacked surface brightness profiles for 304 disc galaxies from
        \citet{Courteau1996,Courteau1997}.
        The profiles have been rescaled using two fit parameters (normalization and \rexp), determined by fitting to a Freeman disc.
        The contours enclose 30, 50, 70 and 90\% of the plotted data points.
        The red line represents the Freeman disc.
        The agreement is similar to that of the fit to the angular momentum distribution in Fig. \ref{fig:ang_mom_hist}.
    }
    \label{fig:sb_stack}
\end{figure}

\begin{figure}
    \includegraphics[width=\columnwidth]{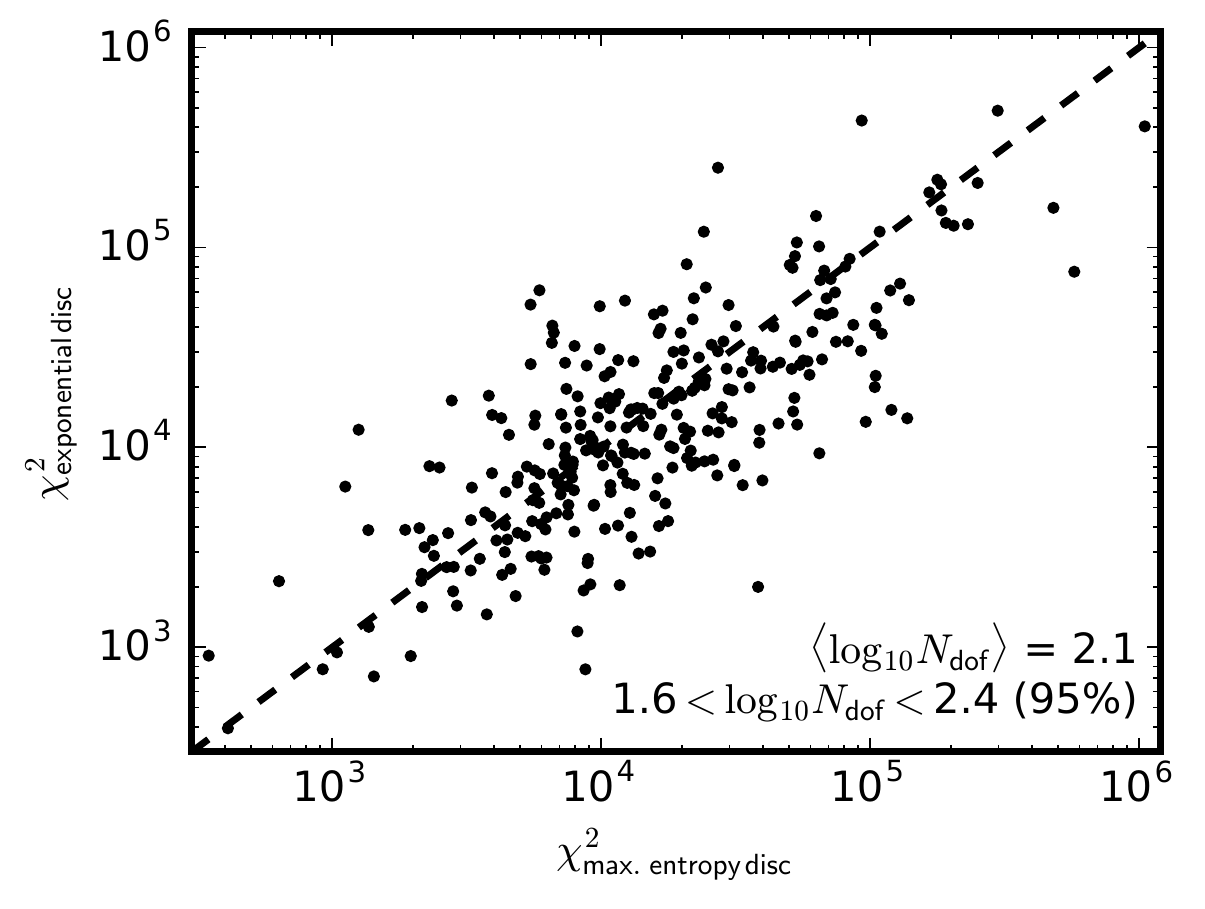}
    \caption{Scatter plot of \chitwo values for fits to a Freeman disc (vertical axis) and the
        maximum entropy model (horizontal axis).
        The dashed line represents the locus of equal \chitwo values for both models.
        \chitwo values have been computed using the inverse of the error of the surface
        brightness data as weights for each data points.
        The average number of degrees of freedom is $\langle \log_{10}N_\mathrm{dof}\rangle=2.1$,
        and 95\% of the galaxies satisfy $1.6<\log_{10}N_\mathrm{dof}<2.4$.
        The maximum entropy model and the Freeman model fit the data about equally well.
        However, the \chitwo values are generally much larger than the number of degrees of
        freedom, mostly because of two properties of the surface brightness profiles of
        galaxy discs: first, azimuthally averaged profiles often have bumps and wiggles due
        to spiral structure or other perturbations that are highly significant in a formal sense;
        second, profiles often show breaks at small and large radii that cannot
        be captured in the functional forms of either model
        (see Fig. \ref{fig:courteau_fit}).
        Clearly, the processes that set the disc profiles have aspects that are too
        complex to be captured by the simple maximum entropy model.
    }
    \label{fig:compare_chi2_scatter}
\end{figure}

For comparison, in Fig. \ref{fig:sb_stack} we show the stacked surface brightness profiles from the galaxies in 
\citet{Courteau1996,Courteau1997}, rescaled by a fit to the usual Freeman disc,
$\Sigma(R)=\Sigma_0\exp(-R/\rexp)$.
The overall agreement, and the scatter, are very similar to that of the fit to the angular momentum distribution
in figure \ref{fig:ang_mom_hist}. 
A more quantitative comparison in Fig. \ref{fig:compare_chi2_scatter} confirms this result.

A different approach is to use the rotation curve from \citet{Courteau1997} to calculate the expected stellar surface density
according to equation \eqref{eq:surface_density}.
This defines a function with two free parameters, \jmean\ and
a normalization constant, which can be fitted to the observed surface brightness data from \citet{Courteau1996}.
We use standard $\chi^2$-minimization with data points weighted by the inverse error in the surface brightness data\footnote{This procedure ignores errors in the rotation curve data and fits.}; the innermost data point at $R=0$ is not included in the fit. 

In Fig. \ref{fig:courteau_fit} we show the fits for fifteen galaxies
randomly selected from the full sample.
The top panel for each galaxy contains the rotation-curve data (orange points and error bars)
along with the Model 2 fit (black line) from \citet{Courteau1997}. The bottom panel shows
the corresponding surface brightness data with the best-fit maximum entropy profile, using the same colour scheme. A few of the surface brightness profiles are fitted remarkably well (e.g., UGC 10536), and most of the profiles are fit well in the central regions.
However, many of the profiles fall below the maximum entropy fit at large radii, possibly because migration operates slowly at large radii and the disc has not had time to approach a maximum entropy state. For about one in four of the galaxies, both in Fig. \ref{fig:courteau_fit} and in the full sample of 304 galaxies, the profile
in the centre is also not reproduced very well.

We may also compare our model to the Freeman disc profile.
We have fit the parameters $\Sigma_0$ and $\rexp$ for each of the galaxies in the sample,
again assuming the same weights (inverse error) per $\log\Sigma$.
In Fig. \ref{fig:compare_chi2_scatter} we show a scatter plot of the respective $\chi^2$
values for the two fits, which each involve two free parameters, a normalization and a scale.
The figure shows that on average the maximum entropy model and the Freeman model fit the data
about equally well; galaxies that fit one model well tend to fit the other and vice versa
although there is significant scatter. The difference is that the Freeman model is empirical
while the maximum entropy model is motivated by general properties of the dynamics of
disc evolution. 

\subsection{Tests with simulations}
\label{sec:simulation_test}

\begin{figure*}
    \includegraphics[width=.97\textwidth]{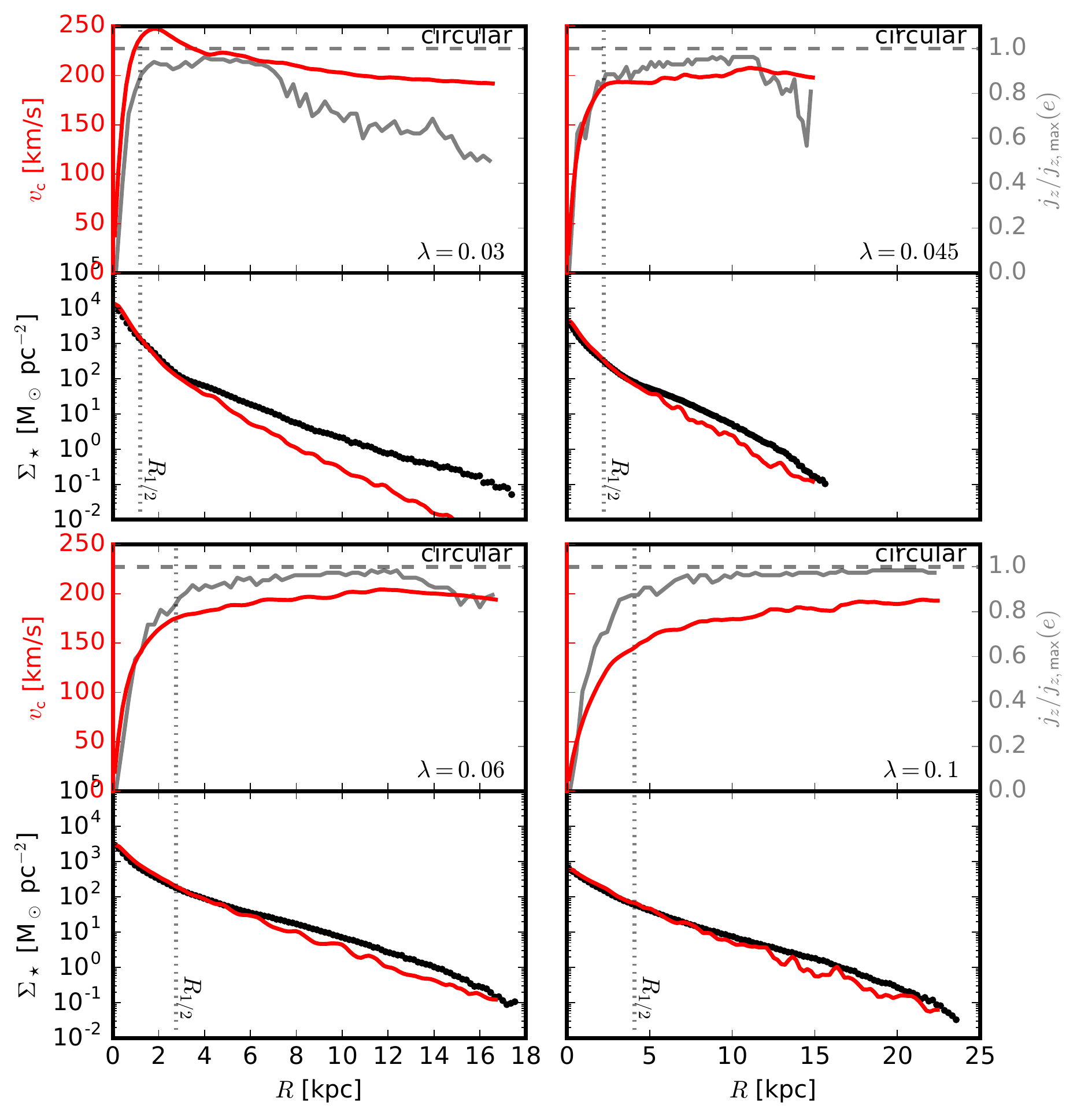}
    \caption{Comparison of maximum entropy discs to the surface density profiles of a subset
        of simulated galaxies from \citet{Herpich2015}.
        For each of the simulated galaxies the upper panel shows the circular-velocity
        curve (red line, left-hand axis) and the circularity parameter (grey line, right-hand
        axis), both as a function of galactocentric radius $R$.
        The corresponding bottom panels show the stellar surface density profile (black points)
        and the maximum entropy profile (red line) based on equation \eqref{eq:surface_density}.
        Note that there are no free parameters although in deriving the maximum entropy model
        we use the known total mass and angular momentum of the disc.
        The wiggles in the predicted surface density profile are due to irregularities in
        the rotation curve that are amplified because the maximum entropy surface density
        depends on the gradient $\d{\vphi}/\d{R}$. The half-mass radius of each simulated galaxy
        is indicated by a vertical dashed line.
        The initial halo spin $\lambda$ of each galaxy is given at the bottom of the top panel;
        in general discs with larger halo spin have larger circularity parameters outside the
        half-mass radius and are better fits to the maximum entropy model, which assumes
        circular orbits.
    }
    \label{fig:sim_test}
\end{figure*}

Simulations of disc galaxy formation offer a controlled way to test whether galaxy discs
evolve towards a maximum entropy state. 
We use the suite of hydrodynamical $N$-body simulations of \citet{Herpich2015} for this purpose. 
These simulations use non-cosmological environments, \ie they lack cosmological
perturbations such as external torques, mergers, fly-bys, etc.
Hence they are well suited to testing our predictions, that are based on the internal
dynamical evolution of the disc (of course, for the same reasons they are not well suited
for exploring whether internal evolution or cosmological perturbations dominate the disc
evolution in real galaxies).
The simulations are fully self-consistent and produce $L^\star$
galaxies with a range of initial halo spin parameters $\lambda$.
A characteristic of this particular suite of simulations is that the simulated galaxies with
low initial halo spin have discs that are dominated by stars on eccentric orbits and those with high initial
spin are dominated by stars on nearly circular  orbits \citep{Herpich2016}.
Thus, only the high-spin simulations satisfy our assumption of nearly circular stellar orbits.
For further details about the simulations see
\citet{Herpich2015}.

For the analysis of the simulations we used the {\sc pynbody} package \citep{pynbody}.
We calculate the circular-velocity or rotation curve in the simulations by differentiating
the gravitational potential $\Phi(R)$ ($\vcirc^2 = R\d{\Phi}/\d{R}$).
We also extract the total stellar mass $M$ and the 
mean specific angular momentum \jmean of each disc from the simulation.
With these data it is straightforward
to predict the surface density profile of the maximum entropy disc and to compare
this to the actual surface density. 
The results are presented in Fig. \ref{fig:sim_test}.
For each galaxy the upper panel shows the circular-velocity curve (red) and the
mode of the circularity parameter $j_z/j_{z, \mathrm{max}}(e)$ (grey)%
\footnote{The circularity parameter is a measure of an orbit's eccentricity.
It compares the angular momentum in $z$-direction to the maximum value possible for a
particle with the same energy in this potential (\ie a particle on a circular orbit).
Thus, it is unity for circular orbits and zero for radial orbits.}.
The corresponding bottom panels show the surface density profile as extracted from the
simulations (black points) and our prediction for a maximum entropy model (red line).
Note that these predictions have no adjustable parameters,
although our procedure guarantees that the total mass and angular momentum of
the maximum entropy disc matches those of the simulated disc.

For almost all galaxies the inner part of the surface density profile (inside the half-mass radius) is reproduced nicely
by our model. 
In the outer part of the disc the agreement gets progressively better with increasing
values for the halo spin parameter $\lambda$; the agreement for the lowest spin galaxy
($\lambda=0.03$) is poor in the outer region, while for the highest spin simulation
($\lambda=0.1$) our prediction follows the actual profile
closely over the entire radial range.
The lower spin simulations might not
represent realistic disc dominated galaxies, as their orbits are far from 
circular even at large radii \citep{Herpich2015}.
We infer that the maximum entropy model provides a remarkably good match to the surface density profiles of simulated disc galaxies with near-circular orbits
(or realistic radial velocity dispersions).

\section{Discussion}
\label{sec:discussion}

\subsection{The radial profile of maximum entropy discs}
\label{sec:summary}

Using arguments from equilibrium statistical mechanics
we have derived an analytic expression for the stellar surface density profiles
of disc galaxies. The profile of these ``maximum entropy" discs depends only on the underlying rotation curve and the total stellar mass and angular
momentum.
The derivation rests on several idealized assumptions:
\begin{itemize}
    \item the disc is axisymmetric, thin, and cold, in the sense that the stellar orbits
        are nearly circular and coplanar;
    \item the total mass and angular momentum of the stars are conserved
        (\ie negligible ongoing star formation, mass loss or accretion, exchange of
        angular momentum with the halo, etc.);
    \item there is a radial mixing mechanism that is efficient enough to distribute the stars uniformly on the phase space manifold allowed by the conserved quantities. 
\end{itemize}

These are strong assumptions that certainly oversimplify the physics of real disc galaxies.
Nevertheless, we have shown in \S\ref{sec:simulation_test} that our model correctly predicts the surface density profiles of
simulated galaxies, at least for isolated discs that are dominated by stars on nearly circular orbits.

We have compared the surface density distribution in maximum entropy discs to the profiles of
304 disc-dominated galaxies from \citet{Courteau1996, Courteau1997}.
This comparison
requires deriving the angular momentum versus radius relation and fitting the model to the
total angular momentum and mass of the observed discs. We found that overall our
maximum entropy model fits about as well as a Freeman disc
\citep{Freeman1970}, an empirical formula chosen to match the observations that has the same
number (2) of free parameters.
Only a few of the galaxies were completely incompatible with maximum entropy models;
the majority of the galaxies show a qualitatively sensible fit; 
and a significant fraction of galaxies yield excellent consistency.

We deem this match encouraging, as it suggests that the physics in our extremely simple
and restrictive model plays 
an important role in determining the radial profiles of galaxy discs.

\subsection{Model limitations}
\label{sec:validity}

It is worth reviewing the simplifying
assumptions and approximations that we made when comparing the model
to observations, which we now do. 
We have assumed that the disc stars are on circular, coplanar orbits, i.e.,
that the disc is cold and razor-thin.
This is a reasonable first approximation: for edge-on disc galaxies the scale height normal
to the disc midplane is approximately independent of radius and only $\sim 10\%$ of the scale length $\rexp$ of
the Freeman model that characterizes the radial distribution of mass in the
disc \citep{Peters2017}. The typical radial velocity 
dispersion $\sigma_R$ in galaxy discs at $\rexp$ is
estimated to be only 30\% of the circular speed $\vphi$ in external galaxies
\citep{vanderKruit_Freeman2011} and the ratio $\sigma_R/\vphi$ is even smaller at larger radii (in the solar
neighbourhood the ratio is $\sim 20\%$). 

However, these approximations are less good close to the centre of the galaxy and the comparison of the maximum entropy model with simulations in \S\ref{sec:simulation_test} suggests
that it breaks down when the assumption of near-circular orbits is not met
(see also Fig. \ref{fig:sim_test}). This is a strong hint that the assumption of circular orbits is necessary.

To explore the effects of relaxing this assumption we may allow the stars to have small but non-zero radial and vertical actions $J_R$ and $J_z$.
Since these are adiabatic invariants they should be conserved by the slow interactions with transient spirals that drive migration \changed{(provided that the influence of the spirals is localized near corotation and does not extend to the Lindblad resonances, as appears to be the case).
This expectation is consistent with $N$-body simulations that show that the vertical and radial actions are approximately conserved even for stars that migrate \citep[\eg][]{Sellwood2002,Minchev2012,Solway2012,vera-ciro_2016}.
If the radial and vertical actions are conserved, the maximum entropy distribution function is $F(j,J_R,J_z)\propto G(J_R,J_z)\exp(-j/\jmean)$ where $G(J_R,J_z)$ is the initial distribution of the disc stars in $J_R$ and $J_z$.}
A more realistic approximation is to include slow growth in $J_R$ and $J_z$ due to scattering by giant molecular clouds and non-corotating spiral arms \changed{\citep{Spitzer1951,Julian1966,Carlberg1987,Fouvry2015}}.
Implementing these generalizations would be straightforward but we have not done so in this paper.
We also note that the migration efficiency decreases as $J_R$ and $J_z$ increase \citep{Sellwood2002,Solway2012,vera-ciro_2016} and it is possible, for example, that the disc is close to a maximum entropy state only for stars on near-circular orbits.

In a strict sense ongoing star formation in discs and infall of halo gas and minor mergers must violate our assumption of total mass and angular momentum conservation. Nonetheless, this approximation may be sensible for our purposes. It is natural to assume that the time-scales for mass ($\tau_M=M_\star/\dot{M_\star}$)
and angular momentum ($\tau_J=J/\dot J$) buildup are similar, $\tau_M\approx\tau_J$.
Then if the migration time $\tau_{\rm mig}\ll \tau_{M, J}$ the disc will always be close to a maximum entropy state although the surface density profile of this state will vary on a time-scale $\tau_{M,J}$.

In a true maximum entropy disc there
should be no radial age or metallicity gradients.
Consequently, there is some tension between our simple model and observations of metallicity and age gradients in disc galaxies.
However, not all disc galaxies have significant gradients and those that do generally 
show only shallow gradients \citep{Sanchez-Blazquez2014,Wilkinson2015,Goddard2016}, which could persist even if the discs were close to maximum entropy. Moreover the youngest stellar populations are the brightest and therefore tend to dominate observed abundance gradients: only stars with ages much longer than the migration time are expected to have a maximum entropy distribution and indeed the metallicity gradient in the Milky Way appears to be smaller for older stars \citep{yu2012}. In general, the maximum entropy disc profile with no gradients can be considered an asymptotic solution, which the discs approach but may not yet have achieved \citep{Schoenrich2017}.

There is strong evidence that substantial radial mixing has occurred in the Milky Way. In the Galaxy, there is a mild, but significant radial abundance gradient among the young stars ($<1$~Gyr), with only a small dispersion in [Fe/H] at any given radius
\citep{Genovali2014,Ness2016}.
But the dispersion in [Fe/H] at any given radius is much larger for stars having ages of a few Gyr
\citep{Ness2016}, arguing for strong radial migration. Additional evidence for migration is 
 the exceptionally low metallicities of some nearby molecular clouds, and the high metallicity of the Sun compared to the average metallicity of the local interstellar medium \citep{Sellwood2002}. Direct evidence for migration is harder to find in other galaxies, although 
\citet{Sanchez-Menguiano2016} found azimuthal variations in chemical abundances in the Sc galaxy GC 6754 that are consistent with large-scale radial migration along the spiral arms \citep{DiMatteo2013,Grand2015,Baba2016,Grand2016}. 

Our arguments so far assume the existence of a radial mixing process that preserves circular, coplanar orbits (i.e., preserves the radial and vertical actions of the orbits).
Radial migration due to transient spirals is an attractive candidate for this process, since it can transport stars radially on time-scales less than the Hubble time and preserves radial actions when the dominant gravitational torques on the stars arise from mass distributions that nearly corotate with them \citep{Sellwood2002,Roskar2012}. We are not aware of other plausible mechanisms in disc galaxies that can shuffle circular, coplanar orbits effectively. 

In general we have computed the distribution of angular momentum or mass in a maximum entropy disc in a given gravitational potential or rotation curve, without asking whether the potential arises from the disc or halo mass distribution, or what determines the halo mass distribution. The only exception is Appendix \ref{sec:self_consistent}, in which we numerically computed the self-consistent surface density profile for an isolated razor-thin disc that is not embedded in a dark halo; however, such a galaxy is not possible in the current standard $\Lambda$ cold dark matter ($\Lambda$CDM) model of cosmology, in which galaxies form from cooling baryons in the potential well of their host halo \citep[\eg][]{Mo2010}. 
In principle both the mass distribution and the rotation curve in maximum entropy discs could be determined following the response of the halo to slow changes in the stellar disc configuration \citep{Blumenthal1986,Gnedin2004}. However, the result would depend on the initial state of the dark halo and the disc and cosmological effects such as gas infall and mergers.
Thus, although the angular momentum distribution in a maximum entropy disc must have the simple form \eqref{eq:dfdefa}, the rotation curve and surface density distribution will depend on the initial conditions and the cosmological context.

In low-mass disc galaxies the validity of our model might be impaired by several effects: (i)  the assumption of circular and coplanar orbits for the stars is less good in dwarf galaxies, because the turbulent velocities in the interstellar medium are a larger fraction of the circular speed; (ii) transient spirals develop most vigorously in strongly shearing discs; low-luminosity galaxies generally have rising rotation curves, and therefore weaker spiral activity than more massive galaxies (J.\ Sellwood, private communication); (iii) 
stellar feedback can alter the gravitational potential and hence the rotation curve by suddenly expelling significant portions of gas on time-scales much less than the migration time  \citep{Navarro1996,Mo2004,Read2005,Mashchenko2006,Maccio2012a,Pontzen2012}; more massive galaxies have deeper gravitational potential wells and are less
susceptible to this process \citep{DiCintio2014}.

\section{Conclusion}
\label{sec:conclusion}

We have described a simple and natural model for the end-state of radial migration in galaxy discsthat leads to an exponential distribution of angular momentum. Given the rotation curve, this model provides a straightforward prediction of the expected surface brightness profile, which fits observed and simulated galaxy discs about as well as the well-established empirical Freeman model (exponential surface brightness). 

Our model is based on the simple ansatz that the angular momenta of disc stars are completely shuffled on time-scales shorter compared to the galaxy age, so the disc entropy is maximized subject to idealized but physically well-motivated constraints. 
The derivation is analogous to that of the Boltzmann distribution for ideal gases. 

Our model stands in a stark contrast to many previous analytical attempts to explain the profiles of galaxy discs, which assume that they are either a consequence of the initial conditions \citep[\eg][]{Fall1980,Dalcanton1997,Dutton2009}, viscous evolution of a gas disc \citep[\eg][]{Lin1987,Yoshii1989}, scattering by interstellar clouds \citep{Elmegreen2013,Elmegreen2016,Struck2017} or a combination of these processes \citep{Ferguson2001}. The assumption of strong migration is consistent with the relatively weak age and metallicity gradients seen in galaxy discs. Of course, discs with even weak population gradients cannot be actually in a maximum entropy state: the scrambling due to migration cannot be complete, but the migration appears to be efficient enough that their surface brightness profiles are fit fairly well by this model. 

\section*{Acknowledgements}
The authors thank Matthias Bartelmann, Walter Dehnen and Jerry Sellwood for enlightening and helpful discussions.
JH and HWR acknowledge funding from the European Research Council under the
European Union's Seventh Framework Programme (FP 7) ERC Advanced Grant Agreement [321035]. ST acknowledges support from NSF grant AST-1406166 and NASA grant NNX14AM24G. HWR acknowledges support of the Miller Institute at the University of California, Berkeley through a visiting professorship  during the completion of this work.

\input{herpich17a.bbl}

\appendix

\section{Energy of maximum entropy discs}
\label{sec:h_theorem}

Here we argue that maximum entropy discs will generally have lower energy than 
other discs with the same mass and angular momentum. 
We model migration by a Fokker--Planck equation of the form
\begin{equation}
\frac{\partial F}{\partial t}= -\frac{\partial}{\partial j}D_1F
+\half\frac{\partial^2}{\partial j^2}D_2F
\end{equation}
where $F(j,t)$ is the time-dependent distribution function and $D_1$
and $D_2\ge 0$ are diffusion coefficients that parametrize the migration
rate. Since migration is due to transient spirals or other mass
concentrations that are much more massive than the stars, the
diffusion coefficients satisfy a fluctuation-dissipation theorem \citep{Binney2008}
\begin{equation}
D_1=\half \frac{\partial D_2}{\partial j}
\end{equation}
so the Fokker--Planck equation simplifies to a diffusion equation of
the form
\begin{equation}
\frac{\partial F}{\partial t}= \half\frac{\partial}{\partial j}D_2\frac{\partial F}{\partial j}.
\end{equation}
This evolution equation conserves the total number of stars if there
is no mass current through the origin or at infinity, $D_2(\partial
F/\partial j)=0$ as $r\to 0$ and $r\to \infty$. The rate of change of total
angular momentum is
\begin{equation}
\dot J =\pi \int \d j\, j\frac{\partial}{\partial j}D_2\frac{\partial
  F}{\partial j}=-\pi\int \d j\, D_2\frac{\partial F}{\partial j},
\end{equation}
where in the last equation we have assumed that there is no
angular momentum current through the origin or infinity. Similarly,
the rate of change of entropy is
\begin{equation}
\dot {\mathfrak S} =\pi\int \d j\, \frac{D_2}{F}\left(\frac{\partial
    F}{\partial j}\right)^2, 
\end{equation}
which is non-negative, as required. If for simplicity we assume that
the disc potential $\Phi(R)$ is fixed, the rate of change of energy is
\begin{equation}
\dot E=2\pi \int \d j\, \Phi(R)\frac{\partial F}{\partial t}= -\pi\int \d
j\,D_2\frac{\partial F}{\partial j} \frac{\d\Phi}{\d R}\frac{\d R}{\d
  j}.
\end{equation}
Let $w(j,t)=D_2(\partial F/\partial j)$ and $s(j)=(\d\Phi/\d R)(\d R/\d
j)$.  Since angular momentum is
conserved we then have
\begin{equation}
\dot J=0=\int \d j\,w(j), \qquad \dot E=-\pi\int \d j\, w(j) s(j).
\end{equation}
If the distribution function is peaked at some intermediate $j$,
then $w(j)$ is positive at
small $j$ and negative at large $j$, 
since the diffusion coefficient $D_2\ge0$. 
For typical galaxy rotation
curves the factor $s(j)$ is positive, but decreases with increasing
$j$---$s(j)\propto j^0$, $j^{-1}$, and $j^{-3}$ for linearly rising,
flat, and Keplerian rotation curves respectively. Consequently, the
integral for $\dot E$ is dominated by small $j$, where $w(j)>0$, so
$\dot E<0$. We conclude that in galaxy discs that do not have pathological rotation curves migration leads to higher entropy and lower energy states.

\section{Self-consistent maximum entropy discs}
\label{sec:self_consistent}

\begin{figure}
    \includegraphics[width=\columnwidth]{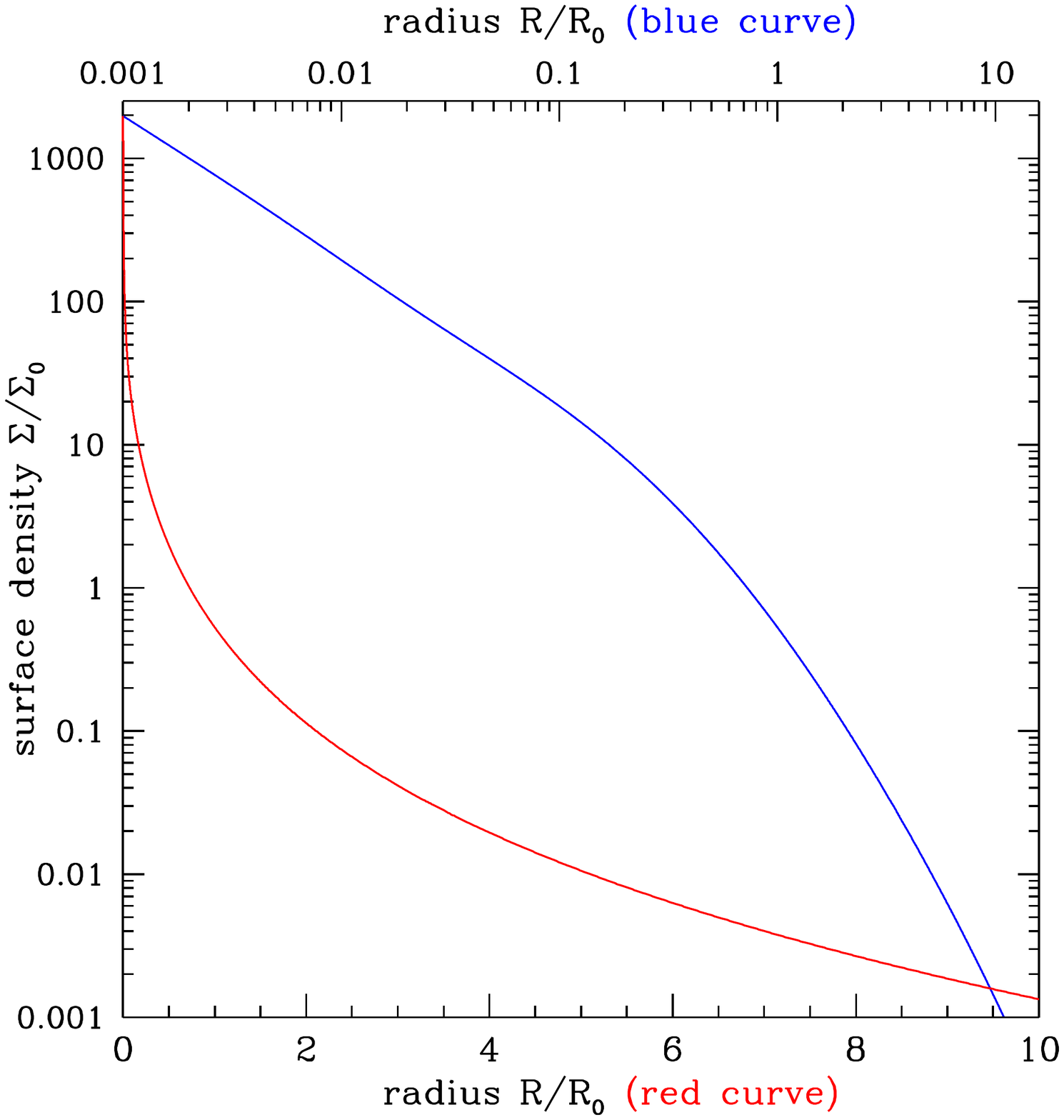}
\vspace{-2.0cm}
    \caption{
        Surface density profile of a self-gravitating maximum entropy disc. The surface density and radius are plotted in units of $\Sigma_0$ and $R_0$ as defined in equation \eqref{eq:sig0}; the radius scale is linear for the blue curve (bottom horizontal axis) and logarithmic for the red curve (top axis). A fitting function for the surface density is given in equations (\ref{eq:fit1}) and (\ref{eq:fit2}). 
    }
    \label{fig:maxent}
\end{figure}

We consider the simple case of a self-gravitating or
self-consistent maximum entropy disc, in which there is no dark-matter halo, no
spheroidal stellar component (bulge), and no significant gas
mass. In this case the disc surface density and circular speed are
related by equations \eqref{eq:kernel}, \eqref{eq:pot}, \eqref{eq:ang_mom_rot_curve}, \eqref{eq:angmom_radius}, and \eqref{eq:surface_density},
with halo potential $\Phi_h(R)=0$. We have solved these equations numerically for the surface density profile as shown in Fig. \ref{fig:maxent}.  

At large radii, the angular momentum is given by $j^2=GMr$. Then equation (\ref{eq:surface_density}) implies that the surface density at large radii is
\begin{equation}\label{eq:large}
\Sigma(R)=\frac{\Sigma_0R_0^{3/2}}{R^{3/2}} \exp\left(-R^{1/2}/R_0^{1/2}\right) \qquad \mbox{as $R\to\infty$},
\end{equation}
with 
\begin{equation} \label{eq:sig0}
R_0\equiv \frac{\jmean^2}{GNm}, \qquad \Sigma_0\equiv \frac{G^2(Nm)^3}{4\pi\jmean^4}.
\end{equation}

At small radii, where $j \ll \jmean$, the density of stars in angular momentum space, $2\pi F(j)$, is approximately constant. In razor-thin self-gravitating discs in centrifugal equilibrium, this angular momentum density distribution is present in two distinct surface density distributions \citep{mestel1963}: (i) cored discs, in which the surface density $\Sigma(R)\sim\mbox{constant}$ near the origin and the angular speed $\Omega=j/R^2$ is also constant; (ii) Mestel-type discs, in which $\Sigma(R)\sim R^{-1}$ and the circular speed $\vphi=j/R$ is constant. Our numerical solutions only produce Mestel-type discs. In such discs the surface density and circular speed at small radii are related by $\Sigma(R)=\vphi^2/(2\pi GR)$ so the number of stars per unit angular momentum is $2\pi R\Sigma(R)/(m\vphi)=\vphi/(Gm)$, which is to be compared to $N/\jmean$ from equation \eqref{eq:dfdefa} in the limit $j\to 0$. Thus $\vphi=GmN/\jmean$ and
\begin{equation}\label{eq:small}
\Sigma(R)=\frac{2\Sigma_0R_0}{R} \qquad \mbox{as $R\to0$}.
\end{equation}

The asymptotic behaviours (\ref{eq:large}) and (\ref{eq:small}) can be interpolated by the formula
\begin{equation} \label{eq:fit1}
\Sigma(R)=\frac{2\Sigma_0R_0\exp[-(R/R_0)^{1/2}]}{R[1+2(R/R_0)^{1/2}]}\big\{1+f[\ln(R/R_0)]\big\}
\end{equation}
where $f(x)\to0$ as $x\to\pm\infty$. Our numerical solutions can be fit by the empirical formula
\begin{align} \label{eq:fit2}
f(x)=&1.1620\exp(-0.1643 x^2-0.8398x)\nonumber\\
&\quad\times (1+0.4097x+0.04993x^2)
\end{align}
with an rms fractional error of 0.4\% between $x=\ln(0.001)$ and $\ln(20)$. 

\begin{thebibliography}{}
\makeatletter
\relax
\def\mn@urlcharsother{\let\do\@makeother \do\$\do\&\do\#\do\^\do\_\do\%\do\~}
\def\mn@doi{\begingroup\mn@urlcharsother \@ifnextchar [ {\mn@doi@}
  {\mn@doi@[]}}
\def\mn@doi@[#1]#2{\def\@tempa{#1}\ifx\@tempa\@empty \href
  {http://dx.doi.org/#2} {doi:#2}\else \href {http://dx.doi.org/#2} {#1}\fi
  \endgroup}
\def\mn@eprint#1#2{\mn@eprint@#1:#2::\@nil}
\def\mn@eprint@arXiv#1{\href {http://arxiv.org/abs/#1} {{\tt arXiv:#1}}}
\def\mn@eprint@dblp#1{\href {http://dblp.uni-trier.de/rec/bibtex/#1.xml}
  {dblp:#1}}
\def\mn@eprint@#1:#2:#3:#4\@nil{\def\@tempa {#1}\def\@tempb {#2}\def\@tempc
  {#3}\ifx \@tempc \@empty \let \@tempc \@tempb \let \@tempb \@tempa \fi \ifx
  \@tempb \@empty \def\@tempb {arXiv}\fi \@ifundefined
  {mn@eprint@\@tempb}{\@tempb:\@tempc}{\expandafter \expandafter \csname
  mn@eprint@\@tempb\endcsname \expandafter{\@tempc}}}

\bibitem[\protect\citeauthoryear{{Baba}, {Morokuma-Matsui}, {Miyamoto}, {Egusa}
   \& {Kuno}}{{Baba} et~al.}{2016}]{Baba2016}
{Baba} J.,  {Morokuma-Matsui} K.,  {Miyamoto} Y.,  {Egusa} F.,   {Kuno} N.,
  2016, \mn@doi [\mnras] {10.1093/mnras/stw987}, \href
  {http://adsabs.harvard.edu/abs/2016MNRAS.460.2472B} {460, 2472}

\bibitem[\protect\citeauthoryear{{Berrier} \& {Sellwood}}{{Berrier} \&
  {Sellwood}}{2015}]{Berrier2015}
{Berrier} J.~C.,  {Sellwood} J.~A.,  2015, \mn@doi [\apj]
  {10.1088/0004-637X/799/2/213}, \href
  {http://adsabs.harvard.edu/abs/2015ApJ...799..213B} {799, 213}

\bibitem[\protect\citeauthoryear{Binney \& Tremaine}{Binney \&
  Tremaine}{2008}]{Binney2008}
Binney J.,  Tremaine S.,  2008, Galactic Dynamics, 2nd edn.
Princeton University Press, Princeton, NJ

\bibitem[\protect\citeauthoryear{{Blumenthal}, {Faber}, {Flores}  \&
  {Primack}}{{Blumenthal} et~al.}{1986}]{Blumenthal1986}
{Blumenthal} G.~R.,  {Faber} S.~M.,  {Flores} R.,   {Primack} J.~R.,  1986,
  \mn@doi [\apj] {10.1086/163867}, \href
  {http://adsabs.harvard.edu/abs/1986ApJ...301...27B} {301, 27}

\bibitem[\protect\citeauthoryear{{Carlberg}}{{Carlberg}}{1987}]{Carlberg1987}
{Carlberg} R.~G.,  1987, \mn@doi [\apj] {10.1086/165702}, \href
  {http://adsabs.harvard.edu/abs/1987ApJ...322...59C} {322, 59}

\bibitem[\protect\citeauthoryear{{Courteau}}{{Courteau}}{1996}]{Courteau1996}
{Courteau} S.,  1996, \mn@doi [\apjs] {10.1086/192281}, \href
  {http://adsabs.harvard.edu/abs/1996ApJS..103..363C} {103, 363}

\bibitem[\protect\citeauthoryear{{Courteau}}{{Courteau}}{1997}]{Courteau1997}
{Courteau} S.,  1997, \mn@doi [\aj] {10.1086/118656}, \href
  {http://adsabs.harvard.edu/abs/1997AJ....114.2402C} {114, 2402}

\bibitem[\protect\citeauthoryear{{Dalcanton}, {Spergel}  \&
  {Summers}}{{Dalcanton} et~al.}{1997}]{Dalcanton1997}
{Dalcanton} J.~J.,  {Spergel} D.~N.,   {Summers} F.~J.,  1997, \apj, \href
  {http://adsabs.harvard.edu/abs/1997ApJ...482..659D} {482, 659}

\bibitem[\protect\citeauthoryear{{Debattista}, {Mayer}, {Carollo}, {Moore},
  {Wadsley}  \& {Quinn}}{{Debattista} et~al.}{2006}]{Debattista2006}
{Debattista} V.~P.,  {Mayer} L.,  {Carollo} C.~M.,  {Moore} B.,  {Wadsley} J.,
   {Quinn} T.,  2006, \mn@doi [\apj] {10.1086/504147}, \href
  {http://adsabs.harvard.edu/abs/2006ApJ...645..209D} {645, 209}

\bibitem[\protect\citeauthoryear{{Di Cintio}, {Brook}, {Macci{\`o}}, {Stinson},
  {Knebe}, {Dutton}  \& {Wadsley}}{{Di Cintio} et~al.}{2014}]{DiCintio2014}
{Di Cintio} A.,  {Brook} C.~B.,  {Macci{\`o}} A.~V.,  {Stinson} G.~S.,  {Knebe}
  A.,  {Dutton} A.~A.,   {Wadsley} J.,  2014, \mn@doi [\mnras]
  {10.1093/mnras/stt1891}, \href
  {http://adsabs.harvard.edu/abs/2014MNRAS.437..415D} {437, 415}

\bibitem[\protect\citeauthoryear{{Di Matteo}, {Haywood}, {Combes}, {Semelin}
  \& {Snaith}}{{Di Matteo} et~al.}{2013}]{DiMatteo2013}
{Di Matteo} P.,  {Haywood} M.,  {Combes} F.,  {Semelin} B.,   {Snaith} O.~N.,
  2013, \mn@doi [\aap] {10.1051/0004-6361/201220539}, \href
  {http://adsabs.harvard.edu/abs/2013A%26A...553A.102D} {553, A102}

\bibitem[\protect\citeauthoryear{{Dutton}}{{Dutton}}{2009}]{Dutton2009}
{Dutton} A.~A.,  2009, \mn@doi [\mnras] {10.1111/j.1365-2966.2009.14741.x},
  \href {http://adsabs.harvard.edu/abs/2009MNRAS.396..121D} {396, 121}

\bibitem[\protect\citeauthoryear{{Elmegreen} \& {Struck}}{{Elmegreen} \&
  {Struck}}{2013}]{Elmegreen2013}
{Elmegreen} B.~G.,  {Struck} C.,  2013, \mn@doi [\apjl]
  {10.1088/2041-8205/775/2/L35}, \href
  {http://adsabs.harvard.edu/abs/2013ApJ...775L..35E} {775, L35}

\bibitem[\protect\citeauthoryear{{Elmegreen} \& {Struck}}{{Elmegreen} \&
  {Struck}}{2016}]{Elmegreen2016}
{Elmegreen} B.~G.,  {Struck} C.,  2016, \mn@doi [\apj]
  {10.3847/0004-637X/830/2/115}, \href
  {http://adsabs.harvard.edu/abs/2016ApJ...830..115E} {830, 115}

\bibitem[\protect\citeauthoryear{{Erwin}, {Beckman}  \& {Pohlen}}{{Erwin}
  et~al.}{2005}]{Erwin2005}
{Erwin} P.,  {Beckman} J.~E.,   {Pohlen} M.,  2005, \mn@doi [\apjl]
  {10.1086/431739}, \href {http://adsabs.harvard.edu/abs/2005ApJ...626L..81E}
  {626, L81}

\bibitem[\protect\citeauthoryear{{Fall} \& {Efstathiou}}{{Fall} \&
  {Efstathiou}}{1980}]{Fall1980}
{Fall} S.~M.,  {Efstathiou} G.,  1980, \mnras, \href
  {http://adsabs.harvard.edu/abs/1980MNRAS.193..189F} {193, 189}

\bibitem[\protect\citeauthoryear{{Ferguson} \& {Clarke}}{{Ferguson} \&
  {Clarke}}{2001}]{Ferguson2001}
{Ferguson} A.,  {Clarke} C.~J.,  2001, \mn@doi [\mnras]
  {10.1046/j.1365-8711.2001.04501.x}, \href
  {http://adsabs.harvard.edu/abs/2001MNRAS.325..781F} {325, 781}

\bibitem[\protect\citeauthoryear{{Fouvry}, {Pichon}, {Magorrian}  \&
  {Chavanis}}{{Fouvry} et~al.}{2015}]{Fouvry2015}
{Fouvry} J.~B.,  {Pichon} C.,  {Magorrian} J.,   {Chavanis} P.~H.,  2015,
  \mn@doi [\aap] {10.1051/0004-6361/201527052}, \href
  {http://adsabs.harvard.edu/abs/2015A%26A...584A.129F} {584, A129}

\bibitem[\protect\citeauthoryear{{Freeman}}{{Freeman}}{1970}]{Freeman1970}
{Freeman} K.~C.,  1970, \mn@doi [\apj] {10.1086/150474}, \href
  {http://adsabs.harvard.edu/abs/1970ApJ...160..811F} {160, 811}

\bibitem[\protect\citeauthoryear{{Fuchs}}{{Fuchs}}{2004}]{Fuchs2004}
{Fuchs} B.,  2004, \mn@doi [\aap] {10.1051/0004-6361:20040098}, \href
  {http://adsabs.harvard.edu/abs/2004A%26A...419..941F} {419, 941}

\bibitem[\protect\citeauthoryear{{Genovali} et~al.,}{{Genovali}
  et~al.}{2014}]{Genovali2014}
{Genovali} K.,  et~al., 2014, \mn@doi [\aap] {10.1051/0004-6361/201323198},
  \href {http://adsabs.harvard.edu/abs/2014A%26A...566A..37G} {566, A37}

\bibitem[\protect\citeauthoryear{{Gnedin}, {Kravtsov}, {Klypin}  \&
  {Nagai}}{{Gnedin} et~al.}{2004}]{Gnedin2004}
{Gnedin} O.~Y.,  {Kravtsov} A.~V.,  {Klypin} A.~A.,   {Nagai} D.,  2004,
  \mn@doi [\apj] {10.1086/424914}, \href
  {http://adsabs.harvard.edu/abs/2004ApJ...616...16G} {616, 16}

\bibitem[\protect\citeauthoryear{{Goddard} et~al.,}{{Goddard}
  et~al.}{2017}]{Goddard2016}
{Goddard} D.,  et~al., 2017, \mn@doi [\mnras] {10.1093/mnras/stw2719}, \href
  {http://adsabs.harvard.edu/abs/2017MNRAS.465..688G} {465, 688}

\bibitem[\protect\citeauthoryear{{Governato} et~al.,}{{Governato}
  et~al.}{2004}]{Governato2004}
{Governato} F.,  et~al., 2004, \mn@doi [\apj] {10.1086/383516}, \href
  {http://adsabs.harvard.edu/abs/2004ApJ...607..688G} {607, 688}

\bibitem[\protect\citeauthoryear{{Governato}, {Willman}, {Mayer}, {Brooks},
  {Stinson}, {Valenzuela}, {Wadsley}  \& {Quinn}}{{Governato}
  et~al.}{2007}]{Governato2007}
{Governato} F.,  {Willman} B.,  {Mayer} L.,  {Brooks} A.,  {Stinson} G.,
  {Valenzuela} O.,  {Wadsley} J.,   {Quinn} T.,  2007, \mn@doi [\mnras]
  {10.1111/j.1365-2966.2006.11266.x}, \href
  {http://adsabs.harvard.edu/abs/2007MNRAS.374.1479G} {374, 1479}

\bibitem[\protect\citeauthoryear{{Grand}, {Kawata}  \& {Cropper}}{{Grand}
  et~al.}{2015}]{Grand2015}
{Grand} R.,  {Kawata} D.,   {Cropper} M.,  2015, \mn@doi [\mnras]
  {10.1093/mnras/stv016}, \href
  {http://adsabs.harvard.edu/abs/2015MNRAS.447.4018G} {447, 4018}

\bibitem[\protect\citeauthoryear{{Grand} et~al.,}{{Grand}
  et~al.}{2016}]{Grand2016}
{Grand} R.,  et~al., 2016, \mn@doi [\mnras] {10.1093/mnrasl/slw086}, \href
  {http://adsabs.harvard.edu/abs/2016MNRAS.460L..94G} {460, L94}

\bibitem[\protect\citeauthoryear{{Guedes}, {Callegari}, {Madau}  \&
  {Mayer}}{{Guedes} et~al.}{2011}]{Guedes2011}
{Guedes} J.,  {Callegari} S.,  {Madau} P.,   {Mayer} L.,  2011, \mn@doi [\apj]
  {10.1088/0004-637X/742/2/76}, \href
  {http://adsabs.harvard.edu/abs/2011ApJ...742...76G} {742, 76}

\bibitem[\protect\citeauthoryear{{Herpich\citedummy{a}}
  et~al.,}{{Herpich\citedummy{a}} et~al.}{2015}]{Herpich2015}
{Herpich\citedummy{a}} J.,  et~al., 2015, \mn@doi [\mnras]
  {10.1093/mnrasl/slv006}, \href
  {http://adsabs.harvard.edu/abs/2015MNRAS.448L..99H} {448, L99}

\bibitem[\protect\citeauthoryear{{Herpich\citedummy{b}}, {Stinson}, {Rix},
  {Martig}  \& {Dutton}}{{Herpich\citedummy{b}} et~al.}{2015}]{Herpich2016}
{Herpich\citedummy{b}} J.,  {Stinson} G.~S.,  {Rix} H.-W.,  {Martig} M.,
  {Dutton} A.~A.,  2015, preprint, \href
  {http://adsabs.harvard.edu/abs/2015arXiv151104442H} {} (\mn@eprint {arXiv}
  {1511.04442})

\bibitem[\protect\citeauthoryear{{Julian} \& {Toomre}}{{Julian} \&
  {Toomre}}{1966}]{Julian1966}
{Julian} W.~H.,  {Toomre} A.,  1966, \mn@doi [\apj] {10.1086/148957}, \href
  {http://adsabs.harvard.edu/abs/1966ApJ...146..810J} {146, 810}

\bibitem[\protect\citeauthoryear{{Katz}}{{Katz}}{1992}]{Katz1992}
{Katz} N.,  1992, \mn@doi [\apj] {10.1086/171366}, \href
  {http://adsabs.harvard.edu/abs/1992ApJ...391..502K} {391, 502}

\bibitem[\protect\citeauthoryear{{Lin} \& {Pringle}}{{Lin} \&
  {Pringle}}{1987}]{Lin1987}
{Lin} D.,  {Pringle} J.~E.,  1987, \mn@doi [\apjl] {10.1086/184981}, \href
  {http://adsabs.harvard.edu/abs/1987ApJ...320L..87L} {320, L87}

\bibitem[\protect\citeauthoryear{{Macci{\`o}}, {Stinson}, {Brook}, {Wadsley},
  {Couchman}, {Shen}, {Gibson}  \& {Quinn}}{{Macci{\`o}}
  et~al.}{2012}]{Maccio2012a}
{Macci{\`o}} A.~V.,  {Stinson} G.,  {Brook} C.~B.,  {Wadsley} J.,  {Couchman}
  H.,  {Shen} S.,  {Gibson} B.~K.,   {Quinn} T.,  2012, \mn@doi [\apjl]
  {10.1088/2041-8205/744/1/L9}, \href
  {http://adsabs.harvard.edu/abs/2012ApJ...744L...9M} {744, L9}

\bibitem[\protect\citeauthoryear{{Mark}}{{Mark}}{1976}]{Mark1976}
{Mark} J.~W.-K.,  1976, \mn@doi [\apj] {10.1086/154396}, \href
  {http://adsabs.harvard.edu/abs/1976ApJ...206..418M} {206, 418}

\bibitem[\protect\citeauthoryear{{Mashchenko}, {Couchman}  \&
  {Wadsley}}{{Mashchenko} et~al.}{2006}]{Mashchenko2006}
{Mashchenko} S.,  {Couchman} H.,   {Wadsley} J.,  2006, \mn@doi [\nat]
  {10.1038/nature04944}, \href
  {http://adsabs.harvard.edu/abs/2006Natur.442..539M} {442, 539}

\bibitem[\protect\citeauthoryear{{Mestel}}{{Mestel}}{1963}]{mestel1963}
{Mestel} L.,  1963, \mn@doi [\mnras] {10.1093/mnras/126.6.553}, \href
  {http://adsabs.harvard.edu/abs/1963MNRAS.126..553M} {126, 553}

\bibitem[\protect\citeauthoryear{{Minchev}, {Famaey}, {Quillen}, {Di Matteo},
  {Combes}, {Vlaji{\'c}}, {Erwin}  \& {Bland-Hawthorn}}{{Minchev}
  et~al.}{2012}]{Minchev2012}
{Minchev} I.,  {Famaey} B.,  {Quillen} A.~C.,  {Di Matteo} P.,  {Combes} F.,
  {Vlaji{\'c}} M.,  {Erwin} P.,   {Bland-Hawthorn} J.,  2012, \mn@doi [\aap]
  {10.1051/0004-6361/201219198}, \href
  {http://adsabs.harvard.edu/abs/2012A%26A...548A.126M} {548, A126}

\bibitem[\protect\citeauthoryear{{Mo} \& {Mao}}{{Mo} \& {Mao}}{2004}]{Mo2004}
{Mo} H.~J.,  {Mao} S.,  2004, \mn@doi [\mnras]
  {10.1111/j.1365-2966.2004.08114.x}, \href
  {http://adsabs.harvard.edu/abs/2004MNRAS.353..829M} {353, 829}

\bibitem[\protect\citeauthoryear{{Mo}, {van den Bosch}  \& {White}}{{Mo}
  et~al.}{2010}]{Mo2010}
{Mo} H.,  {van den Bosch} F.~C.,   {White} S.,  2010, {Galaxy Formation and
  Evolution}.
Cambridge University Press, Cambridge, UK

\bibitem[\protect\citeauthoryear{{Navarro} \& {White}}{{Navarro} \&
  {White}}{1994}]{Navarro1994}
{Navarro} J.~F.,  {White} S.,  1994, \mn@doi [\mnras]
  {10.1093/mnras/267.2.401}, \href
  {http://adsabs.harvard.edu/abs/1994MNRAS.267..401N} {267, 401}

\bibitem[\protect\citeauthoryear{{Navarro}, {Eke}  \& {Frenk}}{{Navarro}
  et~al.}{1996}]{Navarro1996}
{Navarro} J.~F.,  {Eke} V.~R.,   {Frenk} C.~S.,  1996, \mn@doi [\mnras]
  {10.1093/mnras/283.3.L72}, \href
  {http://adsabs.harvard.edu/abs/1996MNRAS.283L..72N} {283, L72}

\bibitem[\protect\citeauthoryear{{Ness}, {Hogg}, {Rix}, {Martig},
  {Pinsonneault}  \& {Ho}}{{Ness} et~al.}{2016}]{Ness2016}
{Ness} M.,  {Hogg} D.~W.,  {Rix} H.-W.,  {Martig} M.,  {Pinsonneault} M.~H.,
  {Ho} A.~Y.~Q.,  2016, \mn@doi [\apj] {10.3847/0004-637X/823/2/114}, \href
  {http://adsabs.harvard.edu/abs/2016ApJ...823..114N} {823, 114}

\bibitem[\protect\citeauthoryear{{Okamoto}, {Eke}, {Frenk}  \&
  {Jenkins}}{{Okamoto} et~al.}{2005}]{Okamoto2005}
{Okamoto} T.,  {Eke} V.~R.,  {Frenk} C.~S.,   {Jenkins} A.,  2005, \mn@doi
  [\mnras] {10.1111/j.1365-2966.2005.09525.x}, \href
  {http://adsabs.harvard.edu/abs/2005MNRAS.363.1299O} {363, 1299}

\bibitem[\protect\citeauthoryear{{Patterson}}{{Patterson}}{1940}]{Patterson1940}
{Patterson} F.~S.,  1940, Harvard College Observatory Bulletin, \href
  {http://adsabs.harvard.edu/abs/1940BHarO.914....9P} {914, 9}

\bibitem[\protect\citeauthoryear{{Peters}, {de Geyter}, {van der Kruit}  \&
  {Freeman}}{{Peters} et~al.}{2017}]{Peters2017}
{Peters} S.~P.~C.,  {de Geyter} G.,  {van der Kruit} P.~C.,   {Freeman} K.~C.,
  2017, \mn@doi [\mnras] {10.1093/mnras/stw2100}, \href
  {http://adsabs.harvard.edu/abs/2017MNRAS.464...48P} {464, 48}

\bibitem[\protect\citeauthoryear{{Pohlen\citedummy{e}}, {Dettmar},
  {L{\"u}tticke}  \& {Aronica}}{{Pohlen\citedummy{e}}
  et~al.}{2002}]{Pohlen2002}
{Pohlen\citedummy{e}} M.,  {Dettmar} R.-J.,  {L{\"u}tticke} R.,   {Aronica} G.,
   2002, \mn@doi [\aap] {10.1051/0004-6361:20020994}, \href
  {http://adsabs.harvard.edu/abs/2002A%26A...392..807P} {392, 807}

\bibitem[\protect\citeauthoryear{{Pohlen\citedummy{h}} \&
  {Trujillo}}{{Pohlen\citedummy{h}} \& {Trujillo}}{2006}]{Pohlen2006}
{Pohlen\citedummy{h}} M.,  {Trujillo} I.,  2006, \mn@doi [\aap]
  {10.1051/0004-6361:20064883}, \href
  {http://adsabs.harvard.edu/abs/2006A%26A...454..759P} {454, 759}

\bibitem[\protect\citeauthoryear{{Pontzen} \& {Governato}}{{Pontzen} \&
  {Governato}}{2012}]{Pontzen2012}
{Pontzen} A.,  {Governato} F.,  2012, \mn@doi [\mnras]
  {10.1111/j.1365-2966.2012.20571.x}, \href
  {http://adsabs.harvard.edu/abs/2012MNRAS.421.3464P} {421, 3464}

\bibitem[\protect\citeauthoryear{{Pontzen}, {Ro{\v s}kar}, {Stinson}  \&
  {Woods}}{{Pontzen} et~al.}{2013}]{pynbody}
{Pontzen} A.,  {Ro{\v s}kar} R.,  {Stinson} G.,   {Woods} R.,  2013, {pynbody:
  N-Body/SPH analysis for python}, Astrophysics Source Code Library (\mn@eprint
  {ascl} {1305.002})

\bibitem[\protect\citeauthoryear{{Read} \& {Gilmore}}{{Read} \&
  {Gilmore}}{2005}]{Read2005}
{Read} J.~I.,  {Gilmore} G.,  2005, \mn@doi [\mnras]
  {10.1111/j.1365-2966.2004.08424.x}, \href
  {http://adsabs.harvard.edu/abs/2005MNRAS.356..107R} {356, 107}

\bibitem[\protect\citeauthoryear{{Robertson}, {Yoshida}, {Springel}  \&
  {Hernquist}}{{Robertson} et~al.}{2004}]{Robertson2004}
{Robertson} B.,  {Yoshida} N.,  {Springel} V.,   {Hernquist} L.,  2004, \mn@doi
  [\apj] {10.1086/382871}, \href
  {http://adsabs.harvard.edu/abs/2004ApJ...606...32R} {606, 32}

\bibitem[\protect\citeauthoryear{{Ro{\v s}kar}, {Debattista}, {Stinson},
  {Quinn}, {Kaufmann}  \& {Wadsley}}{{Ro{\v s}kar} et~al.}{2008a}]{Roskar2008}
{Ro{\v s}kar} R.,  {Debattista} V.~P.,  {Stinson} G.~S.,  {Quinn} T.~R.,
  {Kaufmann} T.,   {Wadsley} J.,  2008a, \mn@doi [\apjl] {10.1086/586734},
  \href {http://adsabs.harvard.edu/abs/2008ApJ...675L..65R} {675, L65}

\bibitem[\protect\citeauthoryear{{Ro{\v s}kar}, {Debattista}, {Quinn},
  {Stinson}  \& {Wadsley}}{{Ro{\v s}kar} et~al.}{2008b}]{Roskar2008a}
{Ro{\v s}kar} R.,  {Debattista} V.~P.,  {Quinn} T.~R.,  {Stinson} G.~S.,
  {Wadsley} J.,  2008b, \mn@doi [\apjl] {10.1086/592231}, \href
  {http://adsabs.harvard.edu/abs/2008ApJ...684L..79R} {684, L79}

\bibitem[\protect\citeauthoryear{{Ro{\v s}kar}, {Debattista}, {Quinn}  \&
  {Wadsley}}{{Ro{\v s}kar} et~al.}{2012}]{Roskar2012}
{Ro{\v s}kar} R.,  {Debattista} V.~P.,  {Quinn} T.~R.,   {Wadsley} J.,  2012,
  \mn@doi [\mnras] {10.1111/j.1365-2966.2012.21860.x}, \href
  {http://adsabs.harvard.edu/abs/2012MNRAS.426.2089R} {426, 2089}

\bibitem[\protect\citeauthoryear{{S{\'a}nchez-Bl{\'a}zquez}
  et~al.,}{{S{\'a}nchez-Bl{\'a}zquez} et~al.}{2014}]{Sanchez-Blazquez2014}
{S{\'a}nchez-Bl{\'a}zquez} P.,  et~al., 2014, \mn@doi [\aap]
  {10.1051/0004-6361/201423635}, \href
  {http://adsabs.harvard.edu/abs/2014A%26A...570A...6S} {570, A6}

\bibitem[\protect\citeauthoryear{{S{\'a}nchez-Menguiano}
  et~al.,}{{S{\'a}nchez-Menguiano} et~al.}{2016}]{Sanchez-Menguiano2016}
{S{\'a}nchez-Menguiano} L.,  et~al., 2016, \mn@doi [\apjl]
  {10.3847/2041-8205/830/2/L40}, \href
  {http://adsabs.harvard.edu/abs/2016ApJ...830L..40S} {830, L40}

\bibitem[\protect\citeauthoryear{Sch{\"o}nrich \& McMillan}{Sch{\"o}nrich \&
  McMillan}{2017}]{Schoenrich2017}
Sch{\"o}nrich R.,  McMillan P.~J.,  2017, \mn@doi [Monthly Notices of the Royal
  Astronomical Society] {10.1093/mnras/stx093}, 467, 1154

\bibitem[\protect\citeauthoryear{{Sellwood\citedummy{e}} \&
  {Binney}}{{Sellwood\citedummy{e}} \& {Binney}}{2002}]{Sellwood2002}
{Sellwood\citedummy{e}} J.~A.,  {Binney} J.~J.,  2002, \mn@doi [\mnras]
  {10.1046/j.1365-8711.2002.05806.x}, \href
  {http://adsabs.harvard.edu/abs/2002MNRAS.336..785S} {336, 785}

\bibitem[\protect\citeauthoryear{{Sellwood\citedummy{h}}}{{Sellwood\citedummy{h}}}{2014}]{Sellwood2014}
{Sellwood\citedummy{h}} J.~A.,  2014, \mn@doi [Rev. Mod. Phys.]
  {10.1103/RevModPhys.86.1}, \href
  {http://adsabs.harvard.edu/abs/2014RvMP...86....1S} {86, 1}

\bibitem[\protect\citeauthoryear{{Solway}, {Sellwood}  \&
  {Sch{\"o}nrich}}{{Solway} et~al.}{2012}]{Solway2012}
{Solway} M.,  {Sellwood} J.~A.,   {Sch{\"o}nrich} R.,  2012, \mn@doi [\mnras]
  {10.1111/j.1365-2966.2012.20712.x}, \href
  {http://adsabs.harvard.edu/abs/2012MNRAS.422.1363S} {422, 1363}

\bibitem[\protect\citeauthoryear{{Sommer-Larsen}, {Gelato}  \&
  {Vedel}}{{Sommer-Larsen} et~al.}{1999}]{Sommer-Larsen1999}
{Sommer-Larsen} J.,  {Gelato} S.,   {Vedel} H.,  1999, \mn@doi [\apj]
  {10.1086/307374}, \href {http://adsabs.harvard.edu/abs/1999ApJ...519..501S}
  {519, 501}

\bibitem[\protect\citeauthoryear{{Sommer-Larsen}, {G{\"o}tz}  \&
  {Portinari}}{{Sommer-Larsen} et~al.}{2003}]{Sommer-Larsen2003}
{Sommer-Larsen} J.,  {G{\"o}tz} M.,   {Portinari} L.,  2003, \mn@doi [\apj]
  {10.1086/377685}, \href {http://adsabs.harvard.edu/abs/2003ApJ...596...47S}
  {596, 47}

\bibitem[\protect\citeauthoryear{{Spitzer} \& {Schwarzschild}}{{Spitzer} \&
  {Schwarzschild}}{1951}]{Spitzer1951}
{Spitzer} Jr. L.,  {Schwarzschild} M.,  1951, \mn@doi [\apj] {10.1086/145478},
  \href {http://adsabs.harvard.edu/abs/1951ApJ...114..385S} {114, 385}

\bibitem[\protect\citeauthoryear{{Steinmetz} \& {Muller}}{{Steinmetz} \&
  {Muller}}{1995}]{Steinmetz1995}
{Steinmetz} M.,  {Muller} E.,  1995, \mn@doi [\mnras]
  {10.1093/mnras/276.2.549}, \href
  {http://adsabs.harvard.edu/abs/1995MNRAS.276..549S} {276, 549}

\bibitem[\protect\citeauthoryear{{Stinson}, {Brook}, {Macci{\`o}}, {Wadsley},
  {Quinn}  \& {Couchman}}{{Stinson} et~al.}{2013}]{Stinson2013}
{Stinson} G.~S.,  {Brook} C.,  {Macci{\`o}} A.~V.,  {Wadsley} J.,  {Quinn}
  T.~R.,   {Couchman} H.,  2013, \mn@doi [\mnras] {10.1093/mnras/sts028}, \href
  {http://adsabs.harvard.edu/abs/2013MNRAS.428..129S} {428, 129}

\bibitem[\protect\citeauthoryear{{Struck} \& {Elmegreen}}{{Struck} \&
  {Elmegreen}}{2017}]{Struck2017}
{Struck} C.,  {Elmegreen} B.~G.,  2017, \mn@doi [\mnras]
  {10.1093/mnras/stw2462}, \href
  {http://adsabs.harvard.edu/abs/2017MNRAS.464.1482S} {464, 1482}

\bibitem[\protect\citeauthoryear{{Tremaine} \& {Ostriker}}{{Tremaine} \&
  {Ostriker}}{1999}]{TO1999}
{Tremaine} S.,  {Ostriker} J.~P.,  1999, \mn@doi [\mnras]
  {10.1046/j.1365-8711.1999.02558.x}, \href
  {http://adsabs.harvard.edu/abs/1999MNRAS.306..662T} {306, 662}

\bibitem[\protect\citeauthoryear{{Vera-Ciro}, {D'Onghia}  \&
  {Navarro}}{{Vera-Ciro} et~al.}{2016}]{vera-ciro_2016}
{Vera-Ciro} C.,  {D'Onghia} E.,   {Navarro} J.~F.,  2016, \mn@doi [\apj]
  {10.3847/1538-4357/833/1/42}, \href
  {http://adsabs.harvard.edu/abs/2016ApJ...833...42V} {833, 42}

\bibitem[\protect\citeauthoryear{{Wilkinson} et~al.,}{{Wilkinson}
  et~al.}{2015}]{Wilkinson2015}
{Wilkinson} D.~M.,  et~al., 2015, \mn@doi [\mnras] {10.1093/mnras/stv301},
  \href {http://adsabs.harvard.edu/abs/2015MNRAS.449..328W} {449, 328}

\bibitem[\protect\citeauthoryear{{Yoshii} \& {Sommer-Larsen}}{{Yoshii} \&
  {Sommer-Larsen}}{1989}]{Yoshii1989}
{Yoshii} Y.,  {Sommer-Larsen} J.,  1989, \mn@doi [\mnras]
  {10.1093/mnras/236.4.779}, \href
  {http://adsabs.harvard.edu/abs/1989MNRAS.236..779Y} {236, 779}

\bibitem[\protect\citeauthoryear{{Yu}, {Sellwood}, {Pryor}, {Chen}  \&
  {Hou}}{{Yu} et~al.}{2012}]{yu2012}
{Yu} J.,  {Sellwood} J.~A.,  {Pryor} C.,  {Chen} L.,   {Hou} J.,  2012, \mn@doi
  [\apj] {10.1088/0004-637X/754/2/124}, \href
  {http://adsabs.harvard.edu/abs/2012ApJ...754..124Y} {754, 124}

\bibitem[\protect\citeauthoryear{{de Vaucouleurs}}{{de
  Vaucouleurs}}{1959}]{deVaucouleurs1959}
{de Vaucouleurs} G.,  1959, Handbuch der Physik, \href
  {http://adsabs.harvard.edu/abs/1959HDP....53..275D} {53, 275}

\bibitem[\protect\citeauthoryear{{van der Kruit} \& {Freeman}}{{van der Kruit}
  \& {Freeman}}{2011}]{vanderKruit_Freeman2011}
{van der Kruit} P.~C.,  {Freeman} K.~C.,  2011, \mn@doi [\araa]
  {10.1146/annurev-astro-083109-153241}, \href
  {http://adsabs.harvard.edu/abs/2011ARA%26A..49..301V} {49, 301}

\makeatother
\end{thebibliography}
\end{document}